\documentclass[a4paper,12pt]{article}
\usepackage{soul}
\usepackage[usenames,dvipsnames]{color}
\usepackage{jheppub}
\usepackage{esvect}
\usepackage{comment}
\usepackage{amsmath, amssymb, slashed, epsf, color, graphicx, latexsym, bm}
\usepackage{mathrsfs}
\usepackage{tensor}
\usepackage{physics}
\usepackage{epsfig}
\usepackage{graphics}
\usepackage{mathtools}
\usepackage{caption}
\usepackage{float}
\usepackage{bbm}
\usepackage{subcaption}
\usepackage{enumitem}
\usepackage{verbatim}
\usepackage{blindtext}
\usepackage[mathlines]{lineno}

\begin{document}
\title{Asymptotic Symmetry algebra of $\mathcal{N}=8$ Supergravity}
\author[a]{Nabamita Banerjee,}
\author[a]{Tabasum Rahnuma,}
\author[b]{and Ranveer Kumar Singh.}
\affiliation[a]{Indian Institute of Science Education and Research Bhopal,
	Bhopal Bypass, Bhopal 462066, India.}
\affiliation[b]{NHETC and Department of physics and Astronomy, Rutgers University, 126
Frelinghuysen Rd., piscataway NJ 08855, USA}
\emailAdd{nabamita@iiserb.ac.in}
\emailAdd{tabasum19@iiserb.ac.in}
\emailAdd{ranveer.singh@rutgers.edu}

\abstract{ The asymptotic symmetry algebra of $\mathcal{N}=1$ supergravity was recently constructed using the well-known $2$D celestial CFT (CCFT) technique \cite{Fotopoulos:2020bqj}. In this paper, we extend the construction to the maximally supersymmetric four dimensional $\mathcal{N}=8$ supergravity theory in asymptotically flat spacetime and construct the extended asymptotic symmetry algebra, which we call $\mathcal{N}=8$ $\mathfrak{sbms}_4$. We use the celestial CFT technique to find the appropriate currents for extensions of $\mathcal{N}=8$ super-Poincar\'{e} and $\mathrm{SU}(8)_R$ R-symmetry current algebra on the celestial sphere $\mathcal{CS}^2$. We generalise the definition of shadow transformations and show that there is \textit{no} infinite dimensional extension of the global $\mathrm{SU}(8)_R$ algebra in the theory. }
\maketitle
\section{Introduction}
The physical observables of a theory are encoded in the symmetries of that theory. This makes the study of symmetries very important. Furthermore, it has been observed that in gauge and gravity theories, there is  an enhancement of symmetry at the boundaries. For asymptotic boundaries, such enhanced symmetries are known as the asymptotic symmetries. In four dimensions these asymptotic symmetries have been studied for both gauge and gravity theories including the $\mathcal{N}=1$ supergravity \cite{He:2014cra,He:2015zea,Dumitrescu:2015fej, PhysRevLett.105.111103,PhysRev.128.2851,Strominger:2013lka, Fotopoulos:2020bqj}. These symmetries in case of gravity and gauge theories are popularly known as BMS (Bondi-Metzner-Sachs) and large gauge symmetries respectively \cite{Bondi:1962px, Sachs:1962wk, Sachs:1962zza,Strominger:2017zoo,He:2014laa,Barnich:2011mi,Barnich:2013axa,Strominger:2013jfa,Gabai:2016kuf}. These infinite dimensional asymptotic symmetries also have experimental implications as in gauge and gravitaional memory effects which are classical observables \cite{Favata:2010zu,PhysRevLett.67.1486,Zeldovich:1974gvh,Susskind:2015hpa,Pasterski:2015zua}. Moreover, a deeper understanding of these symmetries might help in understanding the black hole microscopics \cite{Carlip:2017xne,Sen:2007qy,Mandal:2010cj,Dabholkar:2004yr,Carlip:2012ff}. This necessitates the computation and analysis of asymptotic symmetries. 
\par
Another implication of the asymptotic symmetries is the soft theorems. It has been shown that soft theorems are the Ward identities of the asymptotic symmetries \cite{Strominger:2013jfa,He:2014laa,Campiglia:2019wxe}. Let us discuss this relation in a bit detail.
In general every symmetry leads to constraints on physical observable such as the scattering amplitudes. Such constraints are known as Ward identities. For the Ward identities of asymptotic symmetries we need to consider the amplitude in the soft limit. The soft limit of an amplitude is defined by taking the momenta of one or more external particles to zero. Quite generally, under soft limit, the amplitude factorises into a universal (soft) factor which contains the divergent part of the amplitude times the amplitude without the soft particle(s) insertions. This factorization is known as soft theorem. In other words, soft theorems are the Ward identities of asymptotic symmetries. The BMS and large gauge symmetries lead to soft graviton and soft photon theorem \cite{Strominger:2017zoo}. 
\par
Another important limit of amplitudes is the collinear limit in which the momenta of two external particles are taken to be collinear. Again the amplitude factorises into a collinear factor containing the divergence times the amplitude with the collinear particles replaced by another particle \cite{Taylor:2017sph}. The collinear limit of amplitude turns into an operator product expansion (OPE) of conformal operators of the Celestial Conformal Field theory (CCFT) \cite{Pasterski:2021raf, Pasterski:2021rjz,Donnay:2020guq,Fan:2019emx} as identical momentum directions corresponds to the same operator insertion points on the celestial sphere\footnote{The celestial sphere is the Riemann sphere on the boundary of the Minskowski space.} (which we denote by $\mathcal{CS}^2$). An interesting fact is that the soft and collinear limits of scattering amplitudes can be used to read off the asymptotic symmetries in the context of CCFT \cite{Donnay:2020guq,Fotopoulos:2019vac}. It turns out that to calculate the asymptotic symmetries of a theory, we need to probe the universal soft and collinear sectors of the scattering amplitudes. This has been used to reproduce the BMS algebra in \cite{Fotopoulos:2019vac} and \cite{Banerjee:2022wht} for pure gravity and large gauge algebra for Einstein Yang-Mills theory \cite{Banerjee:2021uxe}. Recently, it has also been used to compute the $\mathcal{N}=1$ supersymmetric extension of the BMS algebra \cite{Fotopoulos:2020bqj}.  
\\\\
In this paper, we calculate the asymptotic symmetries of the four dimensional maximally supersymmetric $\mathcal{N}=8$ supergravity using the CCFT prescription. In celestial CFT of supergravity, the stress tensor is generated by the shadow transform of the soft graviton operator suitably modified to obtain the correct OPE \cite{Banerjee:2022wht}\footnote{see Section \ref{sec:symgen} for more details}, while the supercurrent is generated by the soft gravitino operator \cite{Fotopoulos:2020bqj}. For $\mathcal{N}>1$, the global symmetry algebra contains an additional R-symmetry and hence na\"{i}vely one would expect that the asymptotic algebra would contain an infinite dimensional extension of the global R-symmetry algebra as well. It was shown in \cite{Banerjee:2022abf} that for $\mathcal{N}=2$, even for the $\mathrm{U}(1)^{\mathcal{N}}$ subgroup of the R-symmetry group $\mathrm{U}(\mathcal{N})$ which only scales the supercharges, such an infinite dimensional extension is mathematically inconsistent. For the present paper,  we study the celestial amplitudes of $\mathcal{N}=8$ supergravity and use the soft and collinear limits calculated in a companion paper \cite{tab} to compute the Ward identities and the OPE of conformal operators in the corresponding CCFT. We then construct the stress tensor and the supercurrents of the theory using the shadow transforms of soft graviton and soft gravitino operators. Since the scalars and graviphotinos do not have soft divergences (see \cite{tab}), we are only left with soft graviphoton operators. The R-symmetry current (if any) can then only be constructed using the soft graviphoton operators. We construct the most general such operator present in the CCFT and show that the operator is trivial by requiring that the modes of this operator extend the $\mathrm{SU}(8)_R$ R-symmetry algebra.

The paper is organised as follows: in Section \ref{sec:notprel} we set up our notations and record some definitions and results about the soft and collinear limits in the CCFT of $\mathcal{N}=8$ supergravity used later in the paper. In Section \ref{sec:symgen} we construct the symmetry currents and compute their OPEs. We also construct the possible R-symmetry currents and show that the requirements of R-symmetry extension make the current trivial. Finally in Section \ref{sec:asympsym} we list the full $\mathcal{N}=8~\mathfrak{sbms}_4$ algebra. We conclude in Section \ref{sec:conc} by summarising our results and emphasizing our future goals of the study. The Appendices contain the OPEs of various conformal operators in the Mellin basis computed from the results in \cite{tab} and a detailed calculation of the OPE of the possible R-symmetry currents. 
\section{Notations and preliminaries}\label{sec:notprel}
In this section, we set up the notations for celestial amplitudes and soft and collinear limits in supergravity. 
\subsection{Celestial amplitudes}
Recall that helicity spinors are left and right handed representations of the Lorentz group $\mathrm{SO}(1,3)\sim\mathrm{SL}(2,\mathbb{C})$. We denote the left and right handed helicity spinors by $\lambda_{\alpha}$ and $\tilde{\lambda}^{\dot{\alpha}}$ respectively. A given null momentum $p^{\mu}$ can be written as a bispinor 
\begin{equation}
    p^{\alpha \dot{\alpha}}=\sigma_{\mu}^{\alpha \dot{\alpha}} p^{\mu}=\left(\begin{array}{cc}
p^{0}+p^{3} & p^{1}-i p^{2} \\
p^{1}+i p^{2} & p^{0}-p^{3}
\end{array}\right)=\lambda^{\alpha} \tilde{\lambda}^{\dot{\alpha}}
\end{equation}
where $\sigma_{\mu}^{\alpha \dot{\alpha}}=(1, \sigma_x,\sigma_y,\sigma_z)$. For real physical momentum, the two spinors are related by complex
conjugation $(\tilde{\lambda}^{\dot{\alpha}})^*=\lambda_{\alpha}$. 
\\\\
We now want to study scattering kinematics on the celestial sphere. 
We use the Bondi coordinates $(u,r,z,\bar{z})$ on the Minkowski space where $(z,\bar{z})$ parametrises the celestial sphere $\mathcal{CS}^2$ at null infinity. The Lorentz group $\mathrm{SL}(2,\mathbb{C})$  acts on  $\mathcal{CS}^2$ as follows:
\[
(z,\bar{z})\longmapsto\left(\frac{az+b}{cz+d},\frac{\bar{a}\bar{z}+\bar{b}}{\bar{c}\bar{z}+\bar{d}}\right),\quad \begin{pmatrix}
a&b\\c&d
\end{pmatrix}\in\mathrm{SL}(2,\mathbb{C}).
\]
A general null momentum vector $p^{\mu}$ can be parametrized as 
\[
p^{\mu}=\omega q^{\mu},\quad q^{\mu}=\frac{1}{2}\left(1+|z|^{2}, z+\bar{z},-i(z-\bar{z}), 1-|z|^{2}\right),
\]
where $q^{\mu}$ is a null vector, $\omega$ is identified with the light cone energy and all the particles momenta are taking to be outgoing. Under the Lorentz group the four momentum transforms as a Lorentz vector $p^{\mu}\mapsto \Lambda^{\mu}_{~\nu}p^{\nu}$. This induces the following transformation of $w$ and $q^{\mu}$ as
$$
\omega \mapsto (c z+d)(\bar{c} \bar{z}+\bar{d}) \omega, \quad q^{\mu} \mapsto q^{\prime \mu}=(c z+d)^{-1}(\bar{c} \bar{z}+\bar{d})^{-1} \Lambda_{~\nu}^{\mu} q^{\nu}.
$$ 
In the bispinor notation, we can write the basic null momentum vector $q^{\mu}$ as 
\begin{equation}
    q^{\alpha \dot{\alpha}}=\sigma_{\mu}^{\alpha \dot{\alpha}} q^{\mu}=\left(\begin{array}{cc}
1 &\;  \bar{z} \\
z &\; z \bar{z}
\end{array}\right)=\left(\begin{array}{l}
1 \\
z
\end{array}\right)\left(\begin{array}{ll}
1 & \bar{z}
\end{array}\right),
\end{equation}
where ${\sigma}^{\mu}\equiv (1,\sigma_x,\sigma_y,\sigma_z)$ are two dimensional identity and Pauli matrices. Further introducing the angle and square notation for the left and right handed momentum spinors, we have:
\begin{equation}
\lambda^{\alpha} \equiv\langle p|^{\alpha}=\sqrt{ \omega}\left(\begin{array}{l}
1 \\
z
\end{array}\right)=\sqrt{\omega}\langle q|^{\alpha}, \quad \tilde{\lambda}^{\dot{\alpha}} \equiv| p]^{\dot{\alpha}}=\sqrt{\omega}\left(\begin{array}{l}
1 \\
\bar{z}
\end{array}\right)=\sqrt{\omega} |q]^{\dot{\alpha}},
\end{equation}
where we we write
\begin{equation}
\langle q |^{\alpha}=\left(\begin{array}{l}
1 \\
z
\end{array}\right), \quad| q]^{\dot{\alpha}}=\left(\begin{array}{c}
1 \\
\bar{z}
\end{array}\right).    
\end{equation}
The inner product of momenta can then be written in terms of the angle and square brackets of the corresponding spinors which are now given by
\begin{equation}
 \langle i j\rangle=-\sqrt{\omega_{i} \omega_{j}} z_{i j}, \quad[i j]= \sqrt{\omega_{i} \omega_{j}} \bar{z}_{i j}, 
 \label{eq:paramominncs2}
\end{equation}
where $z_{ij}=z_i-z_j,\bar{z}_{ij}=\bar{z}_i-\bar{z}_j$.\\\\
We can now Mellin transform the fields in the bulk to get conformal primaries on the celestial sphere. The massless conformal primary of conformal dimension $\Delta$ are given by  
\begin{equation}
    \varphi_{\Delta}^{\pm}\left(X^{\mu}, z, \bar{z}\right)=\int_{0}^{\infty} d \omega \omega^{\Delta-1} e^{\pm i \omega q \cdot X-\epsilon \omega}=\frac{(\mp i)^{\Delta} \Gamma(\Delta)}{(-q \cdot X \mp i \epsilon)^{\Delta}}.
\end{equation}
The conformal primaries for nontrivial spins are then given by \cite{Fotopoulos:2020bqj}, 
\begin{equation}
\begin{split}
    & \psi_{\Delta,\ell=-1/2;\alpha}^{\pm}(X,z,\bar{z})=|q\rangle_{\alpha}\varphi^{\pm}_{\Delta+\frac{1}{2}}(X,z,\bar{z})\\& \psi_{\Delta,\ell=1/2}^{\pm;\dot{\alpha}}(X,z,\bar{z})=|q]^{\dot{\alpha}}\varphi^{\pm}_{\Delta+\frac{1}{2}}(X,z,\bar{z})\\&V^{\mu\pm}_{\Delta,\ell=\pm 1}(X,z,\bar{z})=\epsilon^{\mu}_{\ell=\pm 1}(q,r)\varphi^{\pm}_{\Delta}(X,z,\bar{z})\\&
    H_{\Delta, \ell=\pm 2}^{\mu \nu \pm}(X, z, \bar{z}) =\epsilon_{\ell=\pm 1}^{\mu}(q, r) V_{\Delta, \ell=\pm 1}^{\nu \pm}(X, z, \bar{z}) \\ &\psi_{\Delta, \ell=-3 / 2}^{\mu \pm}(X, z, \bar{z}) =\epsilon_{\ell=-1}^{\mu}(q, r) \psi_{\Delta,\ell=-1/2}^{\pm}(X, z, \bar{z})\\&  \bar{\psi}_{\Delta, \ell=+3 / 2}^{\mu \pm}(X, z, \bar{z})=\epsilon_{\ell=+1}^{\mu}(q, r) \bar{\psi}_{\Delta,\ell=+1/2}^{\pm}(X, z, \bar{z}) 
    \label{eq:confprimeltrans}
\end{split}
\end{equation}
where the polarisations are given by 
\begin{equation}
    \epsilon_{\ell=+1}^{\mu}(q, r)=\frac{\left\langle r\left|\sigma^{\mu}\right| q\right]}{\sqrt{2}\langle r q\rangle}, \quad \epsilon_{\ell=-1}^{\mu}(q, r)=\frac{\left[r\left|\bar{\sigma}^{\mu}\right| q\right\rangle}{\sqrt{2}[q r]}
\end{equation}
with $r$ is a reference null vector and $\bar{\sigma}^{\mu}\equiv (1,-\sigma_x,-\sigma_y,-\sigma_z)$. One can further define inner product on these conformal wavepackets \cite{Fotopoulos:2020bqj}. These conformal wavepackets are normalisable only when the conformal dimension $\Delta$ belongs to the principle continuous series, that is $\Delta=1+i\lambda$ with $\lambda\in\mathbb{R}$. In a scattering process, we take all momenta to be outgoing. We now define \textit{celestial amplitude} or \textit{celestial correlator} on $\mathcal{CS}^2$ as the Mellin transform of the amplitudes:
\begin{equation}
    \left\langle\prod_{n=1}^{N} \mathcal{O}_{\Delta_{n}, \ell_{n}}\left(z_{n}, \bar{z}_{n}\right)\right\rangle\equiv\left(\prod_{n=1}^{N} \int d \omega_{n} \omega_{n}^{\Delta_{n}-1}\right) \delta^{(4)}\left(\sum_{n=1}^{N} \omega_{n} q_{n}\right) A_{\ell_{1} \ldots \ell_{N}}\left(\omega_{n}, z_{n}, \bar{z}_{n}\right),
    \label{eq:ccftcor}
\end{equation}
where $A_{\ell_1,\dots,\ell_n}$ is the bulk amplitude with external particles with helicities $\ell_1,\dots,\ell_n$. 
The celestial correlators can be shown to transform as a conformal correlator under $\mathrm{SL}(2,\mathbb{C})$:
\begin{equation}
\left\langle\prod_{n=1}^{N} \mathcal{O}_{\Delta_{n}, \ell_{n}}\left(\frac{az_{n}+b}{cz_n+d}, \frac{\bar{a}\bar{z}_{n}+\bar{b}}{\bar{c}\bar{z}_n+\bar{d}}\right)\right\rangle =\prod_{i=1}^N(cz_i+d)^{\Delta_i+\ell_i}(\bar{c}\bar{z}_i+\bar{d})^{\Delta_i-\ell_i}\left\langle\prod_{n=1}^{N} \mathcal{O}_{\Delta_{n}, \ell_{n}}\left(z_{n}, \bar{z}_{n}\right)\right\rangle.
\label{eq:conftranscelamp}
\end{equation}
where \begin{equation}
    \begin{pmatrix}
        a&b\\c&d
        \end{pmatrix}\in\mathrm{SL}(2,\mathbb{C}).
\end{equation}
\subsection{OPEs of celestial operators in $\mathcal{N}=8$ Supergravity}
Let $\{\eta_A\}_{A=1}^8$ be the Grassmann coordinates on the $\mathcal{N}=8$ superspace. We can package the on-shell degrees of freedom in $\mathcal{N}=8$ supergravity in an on-shell superfield defined as
\begin{equation}
\begin{split}
\Psi(p, \eta)&=H^{+}(p)+\eta_{A} \psi^{A}_+(p)+\eta_{AB} G^{AB}_+(p)+\eta_{ABC} \chi^{ABC}_+(p)\\
&+\eta_{ABCD} \Phi^{ABCD}(p)+\tilde{\eta}^{ABC} \chi_{ABC}^-(p)+\tilde{\eta}^{AB} G_{AB}^-(p)+\tilde{\eta}^{A} \psi_{A}^-(p)+\tilde{\eta} H^{-}(p)
\end{split}
\end{equation}
where we have introduced the notation 
\begin{equation}
\begin{split}
& \eta_{A_{1} \ldots A_{n}} \equiv \frac{1}{n !} \eta_{A_{1}} \ldots \eta_{A_{2}}\\&\tilde{\eta}^{A_{1} \ldots A_{n}} \equiv \epsilon^{A_{1} \ldots A_{n} B_{1} \ldots B_{8-n}} \eta_{B^{1} \ldots B^{8-n}}\\&\tilde{\eta} \equiv \prod_{A=1}^8 \eta^A.   
\end{split}    
\end{equation} 
The fields $H^{\pm}$ represent positive and negative helicity graviton, $G^{AB}_+$ and $G_{AB}^-$ represent positive and negative helicity graviphotons, $\psi^A_+$ and $\psi_A^-$ represent positive and negative helicity gravitinos, $\chi^{ABC}_+$ and $\chi_{ABC}^-$ represent positive and negative helicity graviphotinos and finally $\Phi^{ABCD}$ represent the real scalars. The \textit{superamplitude} is then defined by 
\begin{equation}
  \mathcal{M}_n(\{p_1,\eta^1\},\dots\{p_n,\eta^n\})=\langle\Psi_{1}(p_1,\eta^1)\dots\Psi_{n}(p_n,\eta^n)\rangle.
  \label{eq:supampbulk}  
\end{equation}
This superfield can be Mellin transformed in the usual way to obtain a \textit{celestial superfield} on $\mathcal{CS}^2$, but it turns out that the component fields will have same conformal dimension \cite{Jiang:2021xzy}. This is not appropriate to work with since we want the component fields to have conformal dimensions according to their spin. Thus we work with the so called \textit{quasi-on-shell superfield} \cite{Jiang:2021xzy} defined as
\begin{equation}
    \begin{split}
\Psi_{\Delta}(z,\bar{z},\eta)&=H^+_{\Delta}(z,\bar{z})+\eta_{A} \psi_{\Delta}^A(z,\bar{z})+\eta_{AB} G^{AB}_{\Delta}(z,\bar{z})+\eta_{ABC} \chi^{ABC}_{\Delta}(z,\bar{z}) \\
&+\eta_{ABCD} \Phi^{ABCD}_{\Delta}(z,\bar{z})+\tilde{\eta}^{ABC} \bar{\chi}_{ABC \;\Delta}(z,\bar{z})+\tilde{\eta}^{AB} \bar{G}_{AB \;\Delta}(z,\bar{z})\\
&+\tilde{\eta}^{A} \bar{\psi}_{A\;\Delta}(z,\bar{z})+\tilde{\eta} H^-_{\Delta}(z,\bar{z}),
\end{split}
\end{equation}
where the components are the Mellin transforms of the components fields of $\Psi(p,n)$, all with scaling dimension $\Delta$ as defined in \eqref{eq:confprimeltrans}. The celestial correlator for the component fields can then be defined as in \eqref{eq:ccftcor}. Using the collinear limit of the bulk amplitude, the OPEs of the celestial operators can be computed. To do this computation, we use the collinear limits computed in \cite{tab}. As an example, we calculate the OPE of two graviton operator. The celestial correlator is given by,
\begin{equation}
    \begin{split}
\langle\mathcal{O}_{\Delta_1,+2}\mathcal{O}_{\Delta_2,+2}\dots\mathcal{O}_{\Delta_n,\ell_n}\rangle&=\left(\prod_{j=1}^{n} \int_{0}^{\infty} d \omega_{j} \omega_{j}^{\Delta_j-1}\right) \delta^{4}\left(\sum_{i}\omega_{i} q_{i}\right)M_n\left(1^{+2},2^{+2},\dots, n\right) \\
        &= \left(\prod_{j=3}^{n} \int_{0}^{\infty} d \omega_{j} \omega_{j}^{\Delta_{j}-1} \int_{0}^{\infty} d \omega_{1} \int_{0}^{\infty} d \omega_{2} \omega_{1}^{\Delta_{1}-1} \omega_{2}^{\Delta_{2}-1}\right) \\
& \qquad \times \delta^4\left(\sum_{i=3}^{n} \omega_{i} q_{i}+ \omega_{p} q_{p}\right) \frac{\omega_{p}^{2}}{\omega_{1} \omega_{2}} \frac{\bar{z}_{12}}{z_{12}} M_{n-1}\left(p^{+2}, \ldots, n\right)
\end{split}
\end{equation}
where $M_n$ is the bulk amplitude of component fields and we used the collinear limit 
\begin{equation}
   M_n\left(1^{+2},2^{+2},\dots, n\right) = \frac{\omega_{p}^{2}}{\omega_{1} \omega_{2}} \frac{\bar{z}_{12}}{z_{12}} M_{n-1}\left(p^{+2}, \ldots, n\right).
\end{equation}
Here $p_i=\omega_iq_i,~i=1,2$, the momenta along the collinear channel is $p=p_{1}+p_{2}=\omega_{p} q_{p}$ with $\omega_{p}=\omega_{1}+\omega_{2}$. 
Now we use the following integral \cite{Jiang:2021xzy}:  
\begin{equation}
\int_{0}^{\infty} d \omega_{2} \; \omega_{2}^{\Delta_{2}-1} \int_{0}^{\infty} d \omega_{1}\; \omega_{1}^{\Delta_{1}-1} \; \omega_{1}^{\alpha}\; \omega_{2}^{\beta}\;\omega_p^{\gamma}\; f\left(\omega_p\right)=B\left(\Delta_{1}+\alpha, \Delta_{2}+\beta\right) \int_{0}^{\infty} d \omega_p \;\omega_p^{\Delta_{p}-1} f(\omega_p)
\label{eq:betaintgenrel}
\end{equation}
where $\omega_p=\omega_1+\omega_2$ and $\Delta_p= \Delta_1+\Delta_2+\alpha+\beta+\gamma$ and 
\begin{equation}
    B(x,y)=\frac{\Gamma(x)\Gamma(y)}{\Gamma(x+y)}
\end{equation} 
is the Euler beta function. We get
\begin{equation}
    \begin{split}
\langle\mathcal{O}_{\Delta_1,+2}\mathcal{O}_{\Delta_2,+2}\dots\mathcal{O}_{\Delta_n,\ell_n}\rangle&=\frac{\bar{z}_{12}}{z_{12}}B\left(\Delta_{1}-1,\Delta_{2}-1\right)\left(\prod_{j=3}^{n} \int_{0}^{\infty} d \omega_{j} \omega_{j}^{\Delta_{j}-1} \int_{0}^{\infty} d \omega_{p} \omega_{p}^{ \Delta_{1}+\Delta_{2}-1}\right)\;  \\
& \times \delta^4\left(\sum_{i=3}^{n} \omega_{i} q_{i}+\omega_{p} q_{p}\right)  M_{n-1}\left(p^{+2}, 3, \ldots, n\right) \\
&=\frac{\bar{z}_{12}}{z_{12}} B\left(\Delta_{1}-1, \Delta_{2}-1\right) \;\langle\mathcal{O}_{\Delta_1+\Delta_2,+2}\mathcal{O}_{\Delta_3,\ell_3}\dots\mathcal{O}_{\Delta_n,\ell_n}\rangle
    \end{split}
\end{equation}
This gives the OPE corresponding to the two positive helicity graviton operators,
\begin{equation}
\mathcal{O}_{\Delta_{1},+2}\left(z_{1}, \bar{z}_{1}\right) \mathcal{O}_{\Delta_{2},+2}\left(z_{2,} \bar{z}_{2}\right) \sim \frac{\bar{z}_{12}}{z_{12}} B\left(\Delta_{1}-1, \Delta_{2}-1\right) \mathcal{O}_{\Delta_{1}+\Delta_{2},+2}\left(z_{2}, \bar{z}_{2}\right)
\end{equation}
Similarly for negative helicity gluon we have the collinear amplitude,
\begin{equation}
    \begin{split}
        M_n\left(1^{-2}, 2^{-2},  \ldots\right)&=\frac{\omega_{p}^{2}}{\omega_{1} \omega_{2}} \frac{z_{12}}{\bar{z}_{12}} M_{n-1}\left(p^{-2}, \ldots, n\right)
    \end{split}
\end{equation}
Hence the OPE
\begin{equation}
    \begin{split}
\mathcal{O}_{\Delta_{1},-2}\left(z_{1}, \bar{z}_{1}\right) \mathcal{O}_{\Delta_{2},-2}\left(z_{2}, \bar{z}_{2}\right) \sim \frac{z_{12}}{\bar{z}_{12}} B\left(\Delta_{1}-1 , \Delta_{2}-1\right) \mathcal{O}_{\Delta_{1}+\Delta_{2},+2}\left(z_{2}, \bar{z}_{2}\right)
\end{split}
\end{equation}
The collinear limit of two opposite helicity gravitons is
\begin{equation}
    \begin{split}
M_n\left(1^{+2}, 2^{-2}, 3, \ldots, n\right)=& \frac{\omega_{1}^{3}}{\omega_{p}^{2} \omega_{2}} \frac{\bar{z}_{12}}{z_{12}} M_{n-1}\left(p^{-2}, 3, \ldots, n\right)\\
& +\frac{\omega_{2}^{3}}{\omega_{p}^{2} \omega_{1}} \frac{z_{12}}{\bar{z}_{12}} M_{n-1}\left(p^{+2}, 3, \ldots, n\right)
    \end{split}
\end{equation}
which gives us the OPE
\begin{equation}
    \begin{split}
\mathcal{O}_{\Delta_{1},+2}\left(z_{1}, z_{1}\right) \mathcal{O}_{\Delta_{2},-2}\left(z_{2}, \bar{z}_{2}\right) &=B\left(\Delta_{1}+3, \Delta_{2}-1\right) \frac{\bar{z}_{12}}{z_{12}} \mathcal{O}_{\Delta_{1}+\Delta_{2},-2}\left(z_{2}, \bar{z}_{2}\right) \\
&+B\left(\Delta_{1}-1 , \Delta_{2}+3\right) \frac{z_{12}}{\bar{z}_{12}} \mathcal{O}_{\Delta_{1}+\Delta_{2},+2}\left(z_{2}, \bar{z}_{2}\right) \\
    \end{split}
\end{equation}
One can calculate the OPEs of all other component fields in a similar way using the collinear limits. The results are listed in Appendix \ref{app:opes}.
\subsection{Soft operators in $\mathcal{N}=8$ supergravity}
In the last section we discussed about the collinear limits of amplitudes. In this section we are looking at their soft limits.
As we know, a soft momenta $p\to 0$ can be written as $\omega_p\to 0$ on the celestial sphere and hence an amplitude written in the celestial coordinates can be analysed in the soft limit of any of the external momenta. The result is a soft theorem which expresses an $n$-point amplitude with a soft external momenta $p$ in terms of an $(n-1)$-point amplitude along with a soft factor given by powers of $\omega_p^{-1}$. The various powers of $\omega_p^{-1}$ then correspond to leading, sub-leading, sub-sub-leading soft theorems and so on. Let us first define the \textit{celestial superamplitude} as the Mellin transform of superamplitude:
\begin{equation}
\begin{split}
    \left\langle\prod_{n=1}^{N} \mathcal{O}_{\Delta_{n}}\left(z_{n}, \bar{z}_{n},\eta^n\right)\right\rangle\equiv\left(\prod_{n=1}^{N} \int d \omega_{n} \omega_{n}^{\Delta_{n}-1}\right) \delta^{(4)}\left(\sum_{n=1}^{N} \omega_{n} q_{n}\right)\times\\ \mathcal{M}_N\left(\{\omega_1,z_1,\bar{z}_1,\eta^1\},\dots\{\omega_N,z_N,\bar{z}_N,\eta^N\}\right),
\end{split}
\label{eq:celsupamp}
\end{equation}
where $\mathcal{M}_N\left(\{\omega_1,z_1,\bar{z}_1,\eta^1\},\dots\{\omega_N,z_N,\bar{z}_N,\eta^N\}\right)$ is the superampltude \eqref{eq:supampbulk} written in the celestial basis. We also denote it simply by $\mathcal{M}_N\left(1,2,\dots,N\right)$. 
One can then expand both sides of \eqref{eq:celsupamp} in the Grassmann parameter $\eta_i$ and compare the coefficients to get the celestial amplitude of various component fields. The above expression is identical to that of \eqref{eq:ccftcor}, with the  explicit incorporation of the Grassmann factors in the scattering amplitudes.
\subsubsection{Soft graviton operators}
The leading and sub-leading soft factors corresponding to $\omega_p^{-1}$ and $\omega_p^0$ were calculated in \cite{tab} using double copy relations. Here we only present relevant results and refer the readers to \cite{tab} for further details. In the celestial basis, the leading soft factor is given by
\begin{equation}
    \begin{split}
\mathcal{M}_N(\cdots, j-1, j, j+1, \cdots) \xrightarrow{\omega_j \rightarrow 0} \frac{1}{\omega_j}\sum_{\substack{i=1 \\ i \neq j}}^N \omega_{i}\; z_{j i} \; \bar{z}_{j i} \left[\frac{z^{2}_{j-1, i}}{z^{2}_{j-1, j}\; z_{j, i}^{2}}+\frac{\bar{z}_{j-1, i}^{2}\; }{\bar{z}_{j-1}^{2} \;\bar{z}_{j, i}^{2}}\delta^{4}\left(\eta^{j}\right)\right]\\
\times \mathcal{M}_{N-1}(\cdots, j-1, j+1, \cdots)
\end{split}
\label{eq:leadsoftlims}
\end{equation}
One can now get the soft limit in terms of celestial superamplitude in a straightforward way. We have 
\begin{equation}
    \begin{aligned}
&\left\langle\prod_{n=1}^{N} \lim_{\Delta_j \rightarrow 1} (\Delta_j-1)\mathcal{O}_{\Delta_{n}}\left(z_{n}, \bar{z}_{n}, \eta^n \right)\right\rangle=\left(\prod_{\substack{n=1 \\ n \neq j}}^{N}\int d \omega_{n}\; \omega_{n}^{\Delta_{n}-1}\right)\lim_{\Delta_j\to 1}\int_{0}^{\infty}d\omega_j(\Delta_j-1)\omega_j^{\Delta_j-1} \\&\hspace{5cm}\times\delta^{(4)}\left(\sum_{\substack{k=1 \\ k \neq j}}^{N} \omega_{k} q_{k}\right)
\mathcal{M}_{N}\left(1,\dots,n,\dots,N\right)\\&=\left(\prod_{\substack{n=1 \\ n \neq j}}^{N}\int d \omega_{n}\; \omega_{n}^{\Delta_{n}-1}\right)\int_{0}^{\infty}d\omega_j\frac{d}{d\omega_j}\left(\lim_{\Delta_j\to 1}\omega_j^{\Delta_j-1}\right) \delta^{(4)}\left(\sum_{\substack{k=1 \\ k \neq j}}^{N} \omega_{k} q_{k}\right)
\omega_j\mathcal{M}_{N}\left(1,\dots,n,\dots,N\right)
\end{aligned}
\end{equation}
Using the fact that 
\begin{equation}
  \frac{d}{d\omega_j}\left(\lim_{\Delta_j\to 1}\omega_j^{\Delta_j-1}\right) = \frac{d}{d\omega_j}\theta(\omega_j)=\delta(\omega_j) 
\end{equation}
where $\theta(\omega)$ is the Heaviside step function, we see that the integral on $\omega_j$ on the right hand side gives us 
\begin{equation}
    \begin{aligned}
\left\langle\prod_{n=1}^{N} \lim_{\Delta_j \rightarrow 1} (\Delta_j-1)\mathcal{O}_{\Delta_{n}}\left(z_{n}, \bar{z}_{n}, \eta^n \right)\right\rangle&=\left(\prod_{\substack{n=1 \\ n \neq j}}^{N}\int d \omega_{n}\; \omega_{n}^{\Delta_{n}-1}\right) \delta^{(4)}\left(\sum_{\substack{k=1 \\ k \neq j}}^{N} \omega_{k} q_{k}\right)\\&\hspace{2cm}\times\lim_{\omega_j\to 0}
\omega_j\mathcal{M}_{N}\left(1,\dots,n,\dots,N\right)
\end{aligned}
\end{equation}
Using the soft limit \eqref{eq:leadsoftlims} we get 
\[
\begin{split}
& \left\langle\prod_{n=1}^{N}\lim_{\Delta_j \rightarrow 1} (\Delta_j-1) \mathcal{O}_{\Delta_{n}}\left(z_{n}, \bar{z}_{n}, \eta^n \right)\right\rangle=\sum_{\substack{i=1\\i\neq j}}^{N} \omega_i \Bigg\{\frac{z_{j-1,i}^{2} \bar{z}_{j i}}{z_{j-1, j}^{2} z_{j i}}+\frac{\bar{z}_{j-1,i}^{2} z_{j i}}{\bar{z}_{j-1, j}^{2} \bar{z}_{j i}} \delta^{4}\left(\eta^{j}\right)\Bigg\} \\&\hspace{5cm}\times\left(\prod_{\substack{n=1 \\ n \neq j}}^{N} \int_{0}^{\infty} d \omega_{k} \omega_{k}^{\Delta_{k}-1}\right)\delta^{(4)}\left(\sum_{\substack{k=1 \\ k \neq j}}^{N} \omega_{k} q_{k}\right) \mathcal{M}_{N-1}\left(1,\dots,i,\dots,N\right)\\
&= \sum_{\substack{i=1 \\ i\neq j}}^{n} \Bigg\{\frac{z_{j-1,i}^{2} \bar{z}_{j i}}{z_{j-1, j}^{2} z_{j i}}+\frac{\bar{z}_{j-1,i}^{2} z_{j i}}{\bar{z}_{j-1, j}^{2} \bar{z}_{j i}} \delta^{8}\left(\eta^{j}\right)\Bigg\}\Bigg[\prod_{\substack{n=1 \\ n \neq i}}^{N} \int_{0}^{\infty} d \omega_{k}\; \omega_{k}^{\Delta_{k}-1} \int_{0}^{\infty} d\omega_i \; \omega_i^{\Delta_i} \\&\hspace{9cm}\delta^{(4)}\left(\sum_{\substack{k=1 \\ k \neq j}}^{N} \omega_{k} q_{k}\right) \mathcal{M}_{N-1}\left(1,\dots,i,\dots,N\right)\Bigg]\\
&=\sum_{\substack{i=1 \\ i\neq j}}^{n} \Bigg\{\frac{z_{j-1,i}^{2} \bar{z}_{j i}}{z_{j-1, j}^{2} z_{j i}}+\frac{\bar{z}_{j-1,i}^{2} z_{j i}}{\bar{z}_{j-1, j}^{2} \bar{z}_{j i}} \delta^{8}\left(\eta^{j}\right)\Bigg\}
\left\langle \mathcal{O}_{\Delta_{1}}\left(z_{1}, \bar{z}_{1}, \eta_1 \right), \cdots \mathcal{O}_{\Delta_{i}+1}\left(z_{i}, \bar{z}_{i}, \eta^i \right), \cdots \right\rangle
\end{split}
\]
The Super-Ward identity that we get from the conformally supersoft theorem is
\begin{equation}
\begin{split}
 &\left\langle J_1(z, \bar{z}, \eta) \mathcal{O}_{\Delta_{1}}(z_1, \bar{z}_1, \eta^1) \cdots \mathcal{O}_{\Delta_{N}}\left(z_{N}, \bar{z}_{N}, \eta^N \right)\right\rangle\\
 &=\sum_{i=1}^{N}\left\{\frac{\left(\bar{z}-\bar{z}_{i}\right)}{\left(z-z_{i}\right)} \frac{\left(z_{N}-z_{i}\right)^{2}}{\left(z_{N}-z\right)^{2}}+\frac{\left(z-z_{i}\right)}{\left(\bar{z}-z_{i}\right)} \frac{\left(\bar{z}_{N}-\bar{z}_{i}\right)^{2}}{\left(\bar{z}_{N}-\bar{z}\right)^{2}} \delta^{8}(\eta)\right\} \\
&\hspace{15ex}\times \left\langle \mathcal{O}_{\Delta_{1}}\left(z_{1}, \bar{z}_{1}, \eta^1 \right), \cdots \mathcal{O}_{\Delta_{i}+1}\left(z_{i}, \bar{z}_{i}, \eta^i \right), \cdots, \mathcal{O}_{\Delta_{N}}\left(z_{N}, \bar{z}_{N}, \eta^N \right)\right\rangle 
\end{split}
\end{equation}
where 
\[
J_1(z, \bar{z}, \eta)=\lim_{\Delta\to 1}(\Delta-1)\mathcal{O}_{\Delta}(z,\bar{z},\eta)
\]
is the $\Delta\to 1$ soft operator. 
In the above soft factor we chose the reference vector for polarisation of the soft particle to be the momentum vector of $n$th particle. We leave this reference vector arbitrary which corresponds to a point $\xi\in\mathcal{CS}^2$. The super-Ward identity then takes the form
\begin{equation}
\begin{split}
&\left\langle J_1(z, \bar{z}, \eta) \mathcal{O}_{\Delta_{1}}(z_1, \bar{z}_1, \eta^1) \cdots \mathcal{O}_{\Delta_{N}}\left(z_{N}, \bar{z}_{N}, \eta^N \right)\right\rangle\\ &=\sum_{i=1}^{N}\left\{\frac{\left(\bar{z}-\bar{z}_{i}\right)}{\left(z-z_{i}\right)} \frac{\left(\xi-z_{i}\right)^{2}}{\left(\xi-z\right)^{2}}+\frac{\left(z-z_{i}\right)}{\left(\bar{z}-\bar{z}_{i}\right)} \frac{\left(\bar{\xi}-\bar{z}_{i}\right)^{2}}{\left(\bar{\xi}-\bar{z}\right)^{2}} \delta^{8}(\eta)\right\} \\
&\hspace{15ex}\times \left\langle \mathcal{O}_{\Delta_{1}}\left(z_{1}, \bar{z}_{1}, \eta^1\right), \cdots \mathcal{O}_{\Delta_{i}+1}\left(z_{i}, \bar{z}_{i}, \eta^i \right), \cdots, \mathcal{O}_{\Delta_{N}}\left(z_{N}, \bar{z}_{N}, \eta^N \right)\right\rangle 
\end{split}
\end{equation}
When we expand both sides in the grassmann variables $\eta^i$ and compare coefficients, we get the Ward identity for soft graviton operator:
\begin{equation}
\begin{split}
\left\langle J_1(z,\bar{z})\prod_{n=1}^{N}\mathcal{O}_{\Delta_{n},\ell_n}\left(z_{n}, \bar{z}_{n} \right)\right\rangle=\sum_{i=1}^{N}\frac{\left(\bar{z}-\bar{z}_{i}\right)}{\left(z-z_{i}\right)} &\frac{\left(\xi-z_{i}\right)^{2}}{\left(\xi-z\right)^{2}}\langle \mathcal{O}_{\Delta_{1},\ell_1}\left(z_{1}, \bar{z}_{1}\right),\\& \cdots \mathcal{O}_{\Delta_{i}+1,\ell_i}\left(z_{i}, \bar{z}_{i}\right), \cdots, \mathcal{O}_{\Delta_{N},\ell_N}\left(z_{N}, \bar{z}_{N}\right)\rangle 
\end{split} 
\label{eq:del1gravi}
\end{equation}
and 
\begin{equation}
\begin{split}
\left\langle \bar{J}_1(z,\bar{z})\prod_{n=1}^{N}\mathcal{O}_{\Delta_{n},\ell_n}\left(z_{n}, \bar{z}_{n} \right)\right\rangle=\sum_{i=1}^{N}\frac{\left(z-z_{i}\right)}{\left(\bar{z}-\bar{z}_{i}\right)} &\frac{\left(\bar{\xi}-\bar{z}_{i}\right)^{2}}{\left(\bar{\xi}-\bar{z}\right)^{2}} \langle \mathcal{O}_{\Delta_{1},\ell_1}\left(z_{1}, \bar{z}_{1}\right),\\& \cdots \mathcal{O}_{\Delta_{i}+1,\ell_i}\left(z_{i}, \bar{z}_{i}\right), \cdots, \mathcal{O}_{\Delta_{N},\ell_N}\left(z_{N}, \bar{z}_{N}\right)\rangle 
\end{split} 
\label{eq:del1gravibar}
\end{equation}
where 
\begin{equation}
    J_1(z,\bar{z})=\lim_{\Delta\to 1}(\Delta-1)\mathcal{O}_{\Delta,+2}(z,\bar{z}),\quad \bar{J}_1(z,\bar{z})=\lim_{\Delta\to 1}(\Delta-1)\mathcal{O}_{\Delta,-2}(z,\bar{z})
\end{equation}
are the $\Delta=1$ soft graviton operators.
The subleading soft factor was also calculated in \cite{tab}. It turns out that it is same as the subleading soft factor for positive and negative helicity graviton in pure gravity \cite[Eq. (2.9)]{He:2014bga}. We then write the super Ward identity following the calculations in \cite{Adamo:2019ipt}:
\begin{equation}
    \begin{split}
   &\left\langle J_0(z, \bar{z}, \eta) \mathcal{O}_{\Delta_{1}}(z, \bar{z},\eta^1) \cdots \mathcal{O}_{\Delta_{N}}\left(z_{N}, \bar{z}_{N},\eta^N\right)\right\rangle \\&=\sum_{i=1}^{N}\left\{\frac{\left(\bar{z}-\bar{z}_{i}\right)}{\left(z-z_{i}\right)} \frac{\left(\xi-z_{i}\right)}{\left(\xi-z\right)}((\bar{z}-\bar{z}_i)\partial_{\bar{z}_i}-2\bar{h}_i)+\frac{\left(z-z_{i}\right)}{\left(\bar{z}-\bar{z}_{i}\right)} \frac{\left(\bar{\xi}-\bar{z}_{i}\right)}{\left(\bar{\xi}-\bar{z}\right)} \delta^{8}(\eta)((z-z_i)\partial_{z_i}-2h_i)\right\} \\&\hspace{3cm}\times \left\langle \mathcal{O}_{\Delta_{1}}\left(z_{1}, \bar{z}_{1},\eta^1\right), \cdots  \mathcal{O}_{\Delta_{i}}\left(z_{i}, \bar{z}_{i}, \eta^i \right)\cdots, \mathcal{O}_{\Delta_{N}}\left(z_{N}, \bar{z}_{N},\eta^N\right)\right\rangle 
\end{split}
\end{equation}
where 
\[
J_0(z, \bar{z}, \eta)=\lim_{\Delta\to 0}\Delta[\mathcal{O}_{\Delta,+2}(z, \bar{z})+\delta^8(\eta)\mathcal{O}_{\Delta,-2}(z, \bar{z})].
\]
only contains the $\Delta=0$ soft graviton operators. This immediately gives us the subleading soft graviton limit:
\begin{equation}
\begin{split}
\left\langle J_0(z,\bar{z})\prod_{n=1}^{N}\mathcal{O}_{\Delta_{n},\ell_n}\left(z_{n}, \bar{z}_{n}\right)\right\rangle=\sum_{i=1}^{N}\frac{\left(\bar{z}-\bar{z}_{i}\right)}{\left(z-z_{i}\right)}\frac{\left(\xi-z_{i}\right)}{\left(\xi-z\right)}&((\bar{z}-\bar{z}_i)\partial_{\bar{z}_i}-2\bar{h}_i)\\&\times\left\langle \cdots \mathcal{O}_{\Delta_{i},\ell_i}\left(z_{i}, \bar{z}_{i}\right)\cdots\right\rangle
\end{split} 
\label{eq:del0gravi}
\end{equation}
and 
\begin{equation}
\begin{split}
\left\langle \bar{J}_0(z,\bar{z})\prod_{n=1}^{N}\mathcal{O}_{\Delta_{n},\ell_n}\left(z_{n}, \bar{z}_{n}\right)\right\rangle=\sum_{i=1}^{N}\frac{\left(z-z_{i}\right)}{\left(\bar{z}-\bar{z}_{i}\right)} \frac{\left(\bar{\xi}-\bar{z}_{i}\right)}{\left(\bar{\xi}-\bar{z}\right)}&((z-z_i)\partial_{z_i}-2h_i)\\&\times\left\langle \cdots \mathcal{O}_{\Delta_{i},\ell_i}\left(z_{i}, \bar{z}_{i}\right)\cdots\right\rangle 
\end{split} 
\label{eq:del0gravibar}
\end{equation}
where 
\begin{equation}
    J_0(z,\bar{z})=\lim_{\Delta\to 0}\Delta\mathcal{O}_{\Delta,+2}(z,\bar{z}),\quad \bar{J}_0(z,\bar{z})=\lim_{\Delta\to 0}\Delta\mathcal{O}_{\Delta,-2}(z,\bar{z})
    \label{eq:softGD=0}
\end{equation}
are the $\Delta=0$ soft graviton operators and $h_i=\frac{\Delta_i+\ell_i}{2},\bar{h}_i=\frac{\Delta_i-\ell_i}{2}$ are the conformal weights of the operator $\mathcal{O}_{\Delta_i,\ell_i}(z,\bar{z})$.
\subsubsection{Soft gravitino operator}
Next we move to soft gravitino operator. The leading soft gravitino limit for superamplitudes is given by 
\begin{equation}
\mathcal{M}_{N+1}(\psi_{s+}^A,\{p_1,\eta^1\},\dots\{p_N,\eta^N\})  =\sum_{i=1}^{N} \frac{[si]\langle r i\rangle}{\langle si\rangle\langle r s\rangle}\frac{\partial}{\partial \eta_{iA}} \mathcal{M}_N(\{p_1,\eta^1\},\dots\{p_N,\eta^N\}),  
\end{equation}
where $r$ is the reference vector corresponding to point $\xi\in\mathcal{CS}^2$. The negative helicity soft gravitino limit can be obtained by conjugating the soft factor. We can expand both sides in $\eta^i$ and get the soft theorem in terms of component fields. Note that because of $\partial/\partial_{\eta_{iA}}$, the soft gravitino operator changes the spin of the particle $\ell_i\to\ell_i^c\equiv\ell_i-\frac{1}{2}$. Thus we can only have 
\begin{equation}
\ell_i\in\{-3/2,-1,-1/2,0,+1/2,+1,+3/2,+2\}.
\label{eq:li}
\end{equation}
For negative helicity gravitino $\ell_i^c\to\ell_i$ and clearly 
\begin{equation}
\ell_i^c\in\{-2,-3/2,-1,-1/2,0,+1/2,+1,+3/2\}.  \label{eq:lic} 
\end{equation} 
The explicit soft theorem in terms of celestial amplitudes is given\footnote{we have put the R-symmetry index $*_i$ as a superscript for brevity but it can also be on subscript depending on the helicity of the operator.  Here the number of fermions preceding particle $i$, $\sigma_i=1$ if $\ell_i\in\mathbb{Z}+\frac{1}{2}$ and $0$ otherwise (see \cite{Fotopoulos:2020bqj} for detailed explanation).} by (c.f. \cite[  (6.2),   (6.3)]{Fotopoulos:2020bqj})
\begin{equation}\label{eq:gralim1/2+}
\begin{split}
\left\langle J_{1/2}^A(z,\bar{z})\prod_{n=1}^{N}\mathcal{O}^{*_n}_{\Delta_{n},\ell_n}\left(z_{n}, \bar{z}_{n} \right)\right\rangle=\sum_{i=1}^{N}f(A,\ell_i,*_i,*'_i)&(-1)^{\sigma_i}\frac{\left(\bar{z}-\bar{z}_{i}\right)}{\left(z-z_{i}\right)} \frac{\left(\xi-z_{i}\right)}{\left(\xi-z\right)}\\& \langle \cdots \mathcal{O}^{*'_i}_{\Delta_{i}+\frac{1}{2},\ell_i^c}\left(z_{i}, \bar{z}_{i}\right), \cdots\rangle 
\end{split} 
\end{equation}
and 
\begin{equation}\label{eq:gralim1/2-}
\begin{split}
\left\langle \bar{J}_{1/2~A}(z,\bar{z})\prod_{n=1}^{N}\mathcal{O}^{*_n}_{\Delta_{n},\ell_n^c}\left(z_{n}, \bar{z}_{n} \right)\right\rangle=\sum_{i=1}^{N}\bar{f}(A,\ell_i^c,*_i,*'_i)&(-1)^{\sigma_i}\frac{\left(z-z_{i}\right)}{\left(\bar{z}-\bar{z}_{i}\right)} \frac{\left(\bar{\xi}-\bar{z}_{i}\right)}{\left(\bar{\xi}-\bar{z}\right)} \\& \langle \cdots \mathcal{O}^{*'_i}_{\Delta_{i}+\frac{1}{2},\ell_i}\left(z_{i}, \bar{z}_{i}\right), \cdots\rangle 
\end{split} 
\end{equation}
where 
\begin{equation}
    J^{A}_{1/2}(z,\bar{z})=\lim_{\Delta\to \frac{1}{2}}\left(\Delta-\frac{1}{2}\right)\mathcal{O}^A_{\Delta,+\frac{3}{2}}(z,\bar{z}),\quad \bar{J}_{1/2~A}(z,\bar{z})=\lim_{\Delta\to \frac{1}{2}}\left(\Delta-\frac{1}{2}\right)\mathcal{O}_{\Delta,-\frac{3}{2},A}(z,\bar{z})
    \label{eq:softgravi}
\end{equation}
are soft gravitino operators. The factors $f(A,\ell_i,*_i,*'_i),\bar{f}(A,\ell_i^c,*_i,*'_i)$ are the R-symmetry factors that we can determine using the collinear limits given above. From \eqref{eq:gralim1/2+} and \eqref{eq:gralim1/2-} it is clear that the first argument of $f$ is the R-symmetry index of the soft gravitino operator itself, the second and third arguments are the helicity $\ell_i$ and R-symmetry index $*_i$ respectively of the operator $\mathcal{O}^{*_i}_{\Delta_i, \ell_i}$ which the soft gravitino will act on. Lastly the fourth argument will be the R-symmetry index $*^{'}_i$ of resultant operator. Similarly it goes for $\Bar{f}$. As an example, we can see from the OPE in Eq.\eqref{fermion} that when $\ell_i=-\frac{3}{2}$, $f(A,-3/2,B,*'_i)=\delta^A_B$. Since the resulting particle $\ell=-2$ has no R-symmetry index the $*'_i$ entry is empty.
\\\\
The soft graviphoton limit can be calculated using the OPEs of graviphoton operator with various conformal operators. These OPEs are listed in Appendix \ref{app:opes}. Soft limits  correspond to the values of scaling dimension $\Delta$ of the graviphoton operator for which the beta functions appearing in the OPEs has poles. From Appendix \ref{app:opes} we see that the OPEs of graviphoton operator with various other operators involve\footnote{The OPE of two graviphoton operators with opposite helicity involves another term, c.f. Eq. \eqref{eq:ope+1-1}. One of the terms in the OPE vanishes depending on which of the two helicities of the graviphoton we take to be soft. See Appendix \ref{intco} for such calculations.} $B(\Delta,*)$. Since $B(\Delta,*)$ has poles at all non-positive integer values of $\Delta$, the leading soft limit of the graviphoton operator is $\Delta\to 0$ and all other negative integral values are subleading. In Subsection \ref{sec:R-gencompcalc}, we will need the leading soft graviphoton limit. \par Finally, as noted in \cite{tab}, graviphotino and scalars are trivial in the soft limit and hence do not correspond to any global symmetry \cite{Fotopoulos:2020bqj}. So we do not consider them further. 
\section{Asymptotic Symmetry Generators in $\mathcal{N}=8$ SUGRA}\label{sec:symgen}
Let us first consider the obvious global symmetries of $\mathcal{N}=8$ supergravity. The global symmetry algebra consists of the Poincar\'{e} algebra and the $\mathcal{N}=8$ supersymmetry algebra, together called the $\mathcal{N}=8$ super-Poincar\'{e} algebra and $\mathrm{SU}(8)_R$ $R-$symmetry algebra. At null infinity, we expect to obtain infinite dimensional extensions of these algebras. Following previous works \cite{Fotopoulos:2019vac,Fotopoulos:2020bqj,Banerjee:2021uxe}, we can easily construct the currents that extend the super-Poincar\'{e} algebra, we call this algebra the $\mathcal{N}=8~\mathfrak{sbms}_4$ algebra. We start by constructing the currents for the $\mathcal{N}=8~\mathfrak{sbms}_4$ algebra.  
\subsection{$\mathcal{N}=8~\mathfrak{sbms}_4$ algebra currents}\label{sec:SbarS}
The $\mathfrak{bms}_4$ part of the $\mathcal{N}=8~\mathfrak{sbms}_4$ algebra is known to be generated \cite{Fotopoulos:2019vac} by the shadow transform of the $\Delta=0$ graviton operator suitably modified as discussed below. This is called the generator of superrotations and the level one descendant  of the $\Delta=1$ graviton operator is called the generator of supertranslations on the celestial sphere. Let us define the shadow transforms
$T_0(z,\bar{z})$ and $\overline{T}_0(z,\bar{z})$ as:
\begin{equation}
\begin{aligned}
&T_0(z,\bar{z})=\lim_{\Delta\to 0}\frac{3 !\Delta}{2 \pi} \int d^{2} z^{\prime} \frac{1}{\left(z-z^{\prime}\right)^{4}} \mathcal{O}_{\Delta,-2}\left(z^{\prime}, \bar{z}^{\prime}\right) \\
&\overline{T}_0(z,\bar{z})=\lim_{\Delta\to 0}\frac{3 !\Delta}{2 \pi} \int d^{2} z^{\prime} \frac{1}{\left(\bar{z}-\bar{z}^{\prime}\right)^{4}} \mathcal{O}_{\Delta,+2}\left(z^{\prime}, \bar{z}^{\prime}\right)
\end{aligned}
\label{eq:TTbardef}
\end{equation}
It has been argued in \cite{Banerjee:2022wht} that the above shadow transform operator does not satisfy the usual OPE of a stress tensor. In particular the $T_0T_0$ OPE has an extra term which does not vanish as shown in \cite[Appendix C]{Banerjee:2022wht} unless we modify the stress tensor appropriately. The origin of the problem is the observation that $T_0(z)$ is not holomorphic:
\begin{equation}
    \bar{\partial}T_0=-\frac{1}{2}\partial^3\bar{J}_0(z,\bar{z}) 
\end{equation}
where $J_0$ is the $\Delta=0$ soft graviton operator defined in \eqref{eq:softGD=0}.
Hence the modified the stress tensor can be defined as follows:
\begin{equation}\label{modT}
    T_{\text{mod}}:=T_0+\frac{1}{2}\partial^3\epsilon_{\bar{J}_0}
\end{equation}
where 
\begin{equation}
\epsilon_{\bar{J}_0}:=\int_{\bar{z}_0}^{\bar{z}}d\bar{w}\bar{J}_0(z,\bar{w})     
\end{equation}
with $z_0$ as a reference point. Then it has been shown that the modified stress tensor satisfies the correct $T_{\text{mod}}T_{\text{mod}}$ OPE \cite[Appendix C]{Banerjee:2022wht}. From now on we omit the subscript ``mod" and $T,\overline{T}$ will denote the modified stress tensor.
Using the soft limits \eqref{eq:del1gravi}, \eqref{eq:del1gravibar}, \eqref{eq:del0gravi} and \eqref{eq:del0gravibar} and performing the same calculations as in \cite{Fotopoulos:2019vac}, we arrive at the OPE
\begin{equation}
\begin{split}
    &T(z) \mathcal{O}_{\Delta, \ell}(w, \bar{w})=\frac{h}{(z-w)^{2}} \mathcal{O}_{\Delta, \ell}(w, \bar{w})+\frac{1}{z-w} \partial_{w} \mathcal{O}_{\Delta, \ell}(w, \bar{w})+\text { regular.}\\
    &\overline{T}(\bar{z}) \mathcal{O}_{\Delta, \ell}(w, \bar{w})=\frac{\bar{h}}{(\bar{z}-\bar{w})^{2}} \mathcal{O}_{\Delta, \ell}(w, \bar{w})+\frac{1}{\bar{z}-\bar{w}} \partial_{\bar{w}} \mathcal{O}_{\Delta, \ell}(w, \bar{w})+\operatorname{regular.}
\end{split}
\label{eq:TandO}
\end{equation}
The supertranslations generator $P(z)$,  $\overline{P}(z)$ are defined as:
\begin{equation}
    \begin{aligned}
&P(z)=\lim _{\Delta \rightarrow 1} \frac{(\Delta-1)}{4} \partial_{\bar{z}} \mathcal{O}_{\Delta,+2}(z, \bar{z}) \\
&\overline{P}(\bar{z})=\lim _{\Delta \rightarrow 1} \frac{(\Delta-1)}{4}\partial_{z} \mathcal{O}_{\Delta,-2}(z, \bar{z}) .
\end{aligned}
\label{eq:PPbardef}
\end{equation}
For $P(z)$ we have 
\begin{equation}
 \begin{aligned}
P(z) \mathcal{O}_{\Delta, \ell}(w, \bar{w}) &=\frac{1}{z-w} \mathcal{O}_{\Delta+1, \ell}(w, \bar{w})+\text { regular }
\end{aligned}
\label{eq:PandO}
\end{equation}
and similar OPEs hold for $\overline{P}(\bar{z})$ with conjugated poles. These operators are related to the supertranslation generator $\mathcal{P}(z,\bar{z})$, which is a primary field operator of conformal weight $(\frac{3}{2},\frac{3}{2})$. By contour integrals \cite{Fotopoulos:2019vac}:
\begin{equation}
    P(z)=\frac{1}{2\pi i}\oint d\bar{z}\mathcal{P}(z,\bar{z}),\quad \overline{P}(\bar{z})=\frac{1}{2\pi i}\oint dz\mathcal{P}(z,\bar{z}).
\end{equation}
The supertranslation satisfies the OPE
\begin{equation}
    \mathcal{P}(z,\bar{z})\mathcal{O}_{\Delta,\ell}(w,\bar{w})= \frac{1}{z-w}\frac{1}{\bar{z}-\bar{w}}\mathcal{O}_{\Delta+1,\ell}(w,\bar{w})+\text{regular.}
    \label{eq:curlyPope}
\end{equation}
The supercurrent for $\mathcal{N}=1$ supersymmetry was constructed in \cite{Fotopoulos:2020bqj}. We will see that the same construction will give us the 8 supercurrents for $\mathcal{N}=8$ supersymmetry. We thus define the supercurrents as the shadow transform of the $\Delta=\frac{1}{2}$ gravitino operator:
\begin{equation}\label{Sope}
     \begin{split}
& S_A(z) = \lim_{\Delta \to \frac{1}{2}}\frac{\Delta - \frac{1}{2}}{\pi} \int d^{2} z^{\prime} \frac{1}{\left(z-z^{\prime}\right)^{3}} \mathcal{O}_{A; \Delta,-\frac{3}{2}}(z^{\prime}, \bar{z}^{\prime}) \\
&\overline{S}^{A}(\bar{z}) =\lim_{\Delta \to \frac{1}{2}}\frac{\Delta - \frac{1}{2}}{\pi} \int d^{2} z^{\prime} \frac{1}{\left(\bar{z}-\bar{z}^{\prime}\right)^{3}} \mathcal{O}^A_{\Delta,+\frac{3}{2}}(z^{\prime}, \bar{z}^{\prime})
\end{split}
\end{equation}
Note that the above operators are also not holomorphic since 
\begin{equation}
\begin{split}
\bar{\partial}S_A(z,\Bar{z})= \lim_{\Delta \to 1/2} (\Delta-\frac{1}{2}) \partial_z^2 \mathcal{O}_{A;\Delta, -\frac{3}{2}}(z,\Bar{z})=\partial^2 \Bar{J}_{1/2\; A}(z,\Bar{z})\neq 0
,
\end{split}
\end{equation}
where $\Bar{J}_{1/2\;A}(z,\Bar{z})$ is the leading soft gravitino operator defined in \eqref{eq:softgravi}. One can modify it in a similar way as in Eq. \eqref{modT}. Put 
\begin{equation}
\epsilon_{\Bar{J}_{1/2} \; A}(z,\Bar{z}):=\int_{\Bar{z}_0}^{z} d\Bar{w} \Bar{J}_{1/2\; A}(w,\Bar{w})    
\end{equation}
where $z_0$ is a reference point and define
\begin{equation}
S^A_{\text{mod}}:=S^A-\partial^2\epsilon^{A}_{\Bar{J}_{1/2}}.
\end{equation}
We emphasize that this modification is not required at the quantum level since the OPEs of $S^A$ are as expected for a supercurrent. So we continue to use the shadow transform of the leading soft gravitino operator as the supercurrent without any modification.\\\\
Following the calculations of \cite[Section 7]{Fotopoulos:2020bqj}, it is straightforward to see that
\begin{equation}\label{eq:TSope}
\begin{split}
    &T(z) S_A(w)=\frac{3}{2} \frac{S_A(w)}{(z-w)^2}+\frac{\partial S_A(w)}{z-w}+\text { regular, }\\
&\overline{T}(\bar{z}) \overline{S}^A(\bar{w})=\frac{3}{2} \frac{\overline{S}^A(\bar{z})}{(\bar{z}-\bar{w})^2}+\frac{\bar{\partial} \overline{S}^A(\bar{w})}{\bar{z}-\bar{w}}+\text { regular. }
\end{split}
\end{equation}
and the OPEs $T\overline{S}^A$ and $\overline{T}S_A$ are regular. 
These OPEs confirm the conformal weights of $S_A$ and $\overline{S}_A$ as $(\frac{3}{2},0)$ and $(0,\frac{3}{2})$ respectively.
We now want to show that 
\begin{equation}
    :\{S_B(z),\overline{S}^A(\bar{z})\}:~=~:S_B(z)\overline{S}^A(\bar{z})+\overline{S}^A(\bar{z})S_B(z):~=\delta^A_B\mathcal{P}(z,\bar{z}).
    \label{eq:SSbarnormord}
\end{equation}
Using the gravitino soft limit \eqref{eq:gralim1/2+} and \eqref{eq:gralim1/2-} and the leading graviton limits \eqref{eq:del1gravi} and \eqref{eq:del1gravibar} and following the calculations in \cite[Section 7.3]{Fotopoulos:2020bqj}, we get\footnote{note that we do not seperate the operators in the correlator according to their spins $\ell,\ell^c$  unlike \cite{Fotopoulos:2020bqj} since there is an overlap in the ranges of the two spins. So in the correlators in this calculation, the spins are assumed to be arbitrary.} 
\begin{equation}\label{SbarS}
\begin{split}
    &\left\langle S_B(z)\overline{S}^A(\bar{w})\prod_{n=3}^N\mathcal{O}^{*_n}_{\Delta_n,\ell_n}(z_n,\bar{z}_n)\right\rangle\\&=\delta^A_B\sum_{i=3}^N\left[\frac{1}{(\bar{w}-\bar{z})^2}\frac{\bar{z}-\bar{z}_i}{z-z_i}+\frac{1}{\bar{z}-\bar{w}}\frac{1}{z-z_i}+\frac{1}{\bar{w}-\bar{z}_i}\frac{1}{z-z_i}\right]\left\langle \cdots \mathcal{O}^{*_i}_{\Delta_{i}+1,\ell_i}\left(z_{i}, \bar{z}_{i}\right), \cdots\right\rangle \\&-\sum_{i=3}^Nf(A,\ell_i,*_i,*_i')\bar{f}(B,\ell_i-1/2,*_i',*_i'')\frac{1}{z-z_i}\frac{1}{\bar{w}-\bar{z}_i}\left \langle \cdots \mathcal{O}^{*_i''}_{\Delta_{i}+1,\ell_i}\left(z_{i}, \bar{z}_{i}\right), \cdots\right\rangle \\&+\sum_{\substack{i,j=3\\i\neq j}}^{N}(-1)^{\sigma_i+\sigma_j}f(A,\ell_i,*_i,*_i')\bar{f}(B,\ell_j,*_j,*_j')\frac{1}{z-z_i}\frac{1}{\bar{w}-\bar{z}_j} \left\langle \cdots \mathcal{O}^{*_i'}_{\Delta_{i}+\frac{1}{2},\ell_i-\frac{1}{2}}\left(z_{i}, \bar{z}_{i}\right), \cdots\right.,\\&\hspace{10cm}\left.\mathcal{O}^{*_j'}_{\Delta_{j}+\frac{1}{2},\ell_j+\frac{1}{2}}\left(z_{j}, \bar{z}_{j}\right)\cdots\right\rangle
\end{split}    
\end{equation}
where the factors $f(A,\ell_i,*_i,*_i'),\bar{f}(B,\ell_j,*_j,*_j')$ are the R-symmetry factors that appear on taking the soft or collinear limit depending on the spins and helicities of the soft and collinear particles. In this notation, the first argument of $f$ is the R-symmetry index of the positive helicity soft gravitino, second argument is the spin (and helicity) of one of the hard\footnote{that is not soft} particles, the third argument is the R-symmetry index of that hard particle (left implicit for generality) and the fourth argument is the resulting R-symmetry index of the hard particle after the soft limit is taken (again left implicit for generality). The notation for $\Bar{f}$ is similar.  
It is understood that if the spins do not belong to the required range specified in \eqref{eq:li} and \eqref{eq:lic} then $f,\bar{f}=0$. Similarly 
\begin{equation}\label{barSS}
\begin{split}
    &\left\langle \overline{S}^A(\bar{z})S_B(w)\prod_{n=3}^N\mathcal{O}^{*_n}_{\Delta_n,\ell_n}(z_n,\bar{z}_n)\right\rangle\\&=\delta^A_B\sum_{i=3}^N\left[\frac{1}{(w-z)^2}\frac{z-z_i}{\bar{z}-\bar{z}_i}+\frac{1}{z-w}\frac{1}{\bar{z}-\bar{z}_i}+\frac{1}{\bar{z}-\bar{z}_i}\frac{1}{w-z_i}\right]\left\langle \cdots \mathcal{O}^{*_i}_{\Delta_{i}+1,\ell_i}\left(z_{i}, \bar{z}_{i}\right), \cdots\right\rangle \\&-\sum_{i=3}^N\bar{f}(B,\ell_i,*_i,*_i')f(A,\ell_i+1/2,*_i',*_i'')\frac{1}{w-z_i}\frac{1}{\bar{z}-\bar{z}_i}\left \langle \cdots \mathcal{O}^{*_i''}_{\Delta_{i}+1,\ell_i}\left(z_{i}, \bar{z}_{i}\right), \cdots\right\rangle \\&-\sum_{\substack{i,j=3\\i\neq j}}^{N}(-1)^{\sigma_i+\sigma_j}\bar{f}(B,\ell_i,*_i,*_i')f(A,\ell_j,*_j,*_j')\frac{1}{w-z_i}\frac{1}{\bar{z}-\bar{z}_j} \left\langle \cdots \mathcal{O}^{*_i'}_{\Delta_{i}+\frac{1}{2},\ell_i+\frac{1}{2}}\left(z_{i}, \bar{z}_{i}\right), \cdots\right.,\\&\hspace{10cm}\left.\mathcal{O}^{*_j'}_{\Delta_{j}+\frac{1}{2},\ell_j-\frac{1}{2}}\left(z_{j}, \bar{z}_{j}\right)\cdots\right\rangle
\end{split}    
\end{equation}
Thus the anticommutator is 
\begin{equation}
\begin{split}
    &\left\langle \left(\overline{S}^A(\bar{z})S_B(w)+S_B(z)\overline{S}^A(\bar{w})\right)\prod_{n=3}^N\mathcal{O}^{*_n}_{\Delta_n,\ell_n}(z_n,\bar{z}_n)\right\rangle\\&=\delta^A_B\sum_{i=3}^N\left[\frac{1}{(w-z)^2}\frac{z-z_i}{\bar{z}-\bar{z}_i}+\frac{1}{z-w}\frac{1}{\bar{z}-\bar{z}_i}+\frac{1}{\bar{z}-\bar{z}_i}\frac{1}{w-z_i}\right]\left\langle \cdots \mathcal{O}^{*_i}_{\Delta_{i}+1,\ell_i}\left(z_{i}, \bar{z}_{i}\right), \cdots\right\rangle\\&+\delta^A_B\sum_{i=3}^N\left[\frac{1}{(\bar{w}-\bar{z})^2}\frac{\bar{z}-\bar{z}_i}{z-z_i}+\frac{1}{\bar{z}-\bar{w}}\frac{1}{z-z_i}+\frac{1}{\bar{w}-\bar{z}_i}\frac{1}{z-z_i}\right]\left\langle \cdots \mathcal{O}^{*_i}_{\Delta_{i}+1,\ell_i}\left(z_{i}, \bar{z}_{i}\right), \cdots\right\rangle \\&-\sum_{i=3}^Nf(A,\ell_i,*_i,*_i')\bar{f}(B,\ell_i-1/2,*_i',*_i'')\frac{1}{z-z_i}\frac{1}{\bar{w}-\bar{z}_i}\left \langle \cdots \mathcal{O}^{*_i''}_{\Delta_{i}+1,\ell_i}\left(z_{i}, \bar{z}_{i}\right), \cdots\right\rangle \\&-\sum_{i=3}^N\bar{f}(B,\ell_i,*_i,*_i')f(A,\ell_i+1/2,*_i',*_i'')\frac{1}{w-z_i}\frac{1}{\bar{z}-\bar{z}_i}\left \langle \cdots \mathcal{O}^{*_i''}_{\Delta_{i}+1,\ell_i}\left(z_{i}, \bar{z}_{i}\right), \cdots\right\rangle.
\end{split} 
\end{equation}
Here in the last terms in Eq.\eqref{SbarS} and Eq.\eqref{barSS} we have relative signs hence both terms cancel. One can notice that relative sign is due to the action of $S$ and $\overline{S}$ on different clusters for $ i<j$ and $i>j$ in both the terms. Then we see that the normal ordered current $:\{S_B(z),\overline{S}^A(\bar{z})\}:$ satisfies 
\begin{equation}
\begin{split}
    &\left\langle :\{S_B(z),\overline{S}^A(\bar{z})\}:\prod_{n=3}^N\mathcal{O}^{*_n}_{\Delta_n,\ell_n}(z_n,\bar{z}_n)\right\rangle\\&=2\delta^A_B\sum_{i=3}^N\frac{1}{\bar{z}-\bar{z}_i}\frac{1}{z-z_i}\left\langle \cdots\mathcal{O}^{*_i}_{\Delta_{i}+1,\ell_i}\left(z_{i}, \bar{z}_{i}\right), \cdots\right\rangle\\&-\sum_{i=3}^Nf(A,\ell_i,*_i,*_i')\bar{f}(B,\ell_i-1/2,*_i',*_i'')\frac{1}{z-z_i}\frac{1}{\bar{z}-\bar{z}_i}\left \langle \cdots \mathcal{O}^{*_i''}_{\Delta_{i}+1,\ell_i}\left(z_{i}, \bar{z}_{i}\right), \cdots\right\rangle \\&-\sum_{i=3}^N\bar{f}(B,\ell_i,*_i,*_i')f(A,\ell_i+1/2,*_i',*_i'')\frac{1}{z-z_i}\frac{1}{\bar{z}-\bar{z}_i}\left \langle \cdots \mathcal{O}^{*_i''}_{\Delta_{i}+1,\ell_i}\left(z_{i}, \bar{z}_{i}\right), \cdots\right\rangle.
\end{split} 
\label{eq:ope[ss]barff}
\end{equation}
We now show that for any $\ell_i$, R-symmetry factors in the last two sums reduce to $\delta_B^A$. Let us start with $\ell_i=+2$ in which case $*_i,*_i''$ is empty. Moreover in this case $\bar{f}(B,+2,*_i,*_i')=0$ so that we only have one term to analyse. From the OPEs in Appendix \ref{app:opes}, we see that $*_i'=A$ and
\begin{equation}
 f(A,+2,{}_{-},*_i')\bar{f}(B,+3/2,*_i',{}_{-})\mathcal{O}_{\Delta_{i}+1,+2}\left(z_{i}, \bar{z}_{i}\right)= \delta^A_B \mathcal{O}_{\Delta_{i}+1,+2}\left(z_{i}, \bar{z}_{i}\right).
\end{equation}
The case $\ell_i=+\frac{3}{2}$ is more interesting. Suppose $*_i=C$ then from the OPEs, we can easily see that $*_i'=AC$ for the second term and $*_i'$ is empty for the last term. We then have 
\begin{equation}
f(A,+3/2,C,*_i')\bar{f}(B,+1,*_i',*_i'')\mathcal{O}^{*_i''}_{\Delta_{i}+1,+3/2}\left(z_{i}, \bar{z}_{i}\right)=2!\delta^{[A}_B\mathcal{O}^{C]}_{\Delta_{i}+1,+3/2}\left(z_{i}, \bar{z}_{i}\right)   
\end{equation}
and similarly
\begin{equation}
\bar{f}(B,+3/2,C,{}_{-})f(A,+2,{}_{-},*_i'')\mathcal{O}^{*_i''}_{\Delta_{i}+1,+3/2}\left(z_{i}, \bar{z}_{i}\right)=\delta^C_B\mathcal{O}^{A}_{\Delta_{i}+1,+3/2}\left(z_{i}, \bar{z}_{i}\right)     
\end{equation}
We can clearly see that the sum of the last two terms is simply $\delta_{B}^A\mathcal{O}^{C}_{\Delta_{i}+1,+3/2}\left(z_{i}, \bar{z}_{i}\right)$. The case $\ell_i=-\frac{3}{2}$ is similar. Let us now analyse the case $\ell_i=+1$ in which case $*_i=CD$. We get 
\begin{equation}
\begin{split}
f(A,+1,CD,*_i')\bar{f}(B,+1/2,*_i',*_i'')&\mathcal{O}^{*_i''}_{\Delta_{i}+1,+2}\left(z_{i}, \bar{z}_{i}\right)\\&=  \bar{f}(B,+1/2,*_i',ACD)\mathcal{O}^{ACD}_{\Delta_{i}+\frac{1}{2},+1/2}\left(z_{i}, \bar{z}_{i}\right)\\&=3\delta^{[A}_{B}\mathcal{O}^{CD]}_{\Delta_{i}+1,+1}\left(z_{i}, \bar{z}_{i}\right). 
\end{split}
\end{equation}
Similarly
\begin{equation}
\bar{f}(B,+1,CD,*_i')f(A,+3/2,*_i',*_i'')\mathcal{O}^{*_i''}_{\Delta_{i}+1,+3/2}\left(z_{i}, \bar{z}_{i}\right)=-2!\delta^{[C}_{B}\mathcal{O}^{D]A}_{\Delta_{i}+1,+1}\left(z_{i}, \bar{z}_{i}\right)       
\end{equation}
which finally implies 
\begin{equation}
\begin{split}
3\delta^{[A}_{B}\mathcal{O}^{CD]}- 2!\delta^{[C}_{B}\mathcal{O}^{D]A}&=\frac{1}{2}\left[(\delta^{A}_{B}\mathcal{O}^{CD}-\delta^{A}_{B}\mathcal{O}^{DC})+(\delta^{C}_{B}\mathcal{O}^{DA}-\delta^{C}_{B}\mathcal{O}^{AD})+(\delta^{D}_{B}\mathcal{O}^{AC}-\delta^{D}_{B}\mathcal{O}^{CA})\right]\\&-\delta^{C}_{B}\mathcal{O}^{DA}+\delta^{D}_{B}\mathcal{O}^{CA}\\&=\delta^{A}_{B}\mathcal{O}^{CD}
\end{split}
\end{equation}
The case $\ell_i=-1$ is similar. The same calculation as in $\ell_i=1$ recurs for the cases $\ell_i=1/2,0$. \\\\
These calculations simplify the OPE \eqref{eq:ope[ss]barff}. We get 
\begin{equation}
\left\langle :\{S_B(z),\overline{S}^A(\bar{z})\}:\prod_{n=3}^N\mathcal{O}^{*_n}_{\Delta_n,\ell_n}(z_n,\bar{z}_n)\right\rangle=\delta^A_B\sum_{i=3}^N\frac{1}{\bar{z}-\bar{z}_i}\frac{1}{z-z_i}\left\langle \cdots\mathcal{O}^{*_i}_{\Delta_{i}+1,\ell_i}\left(z_{i}, \bar{z}_{i}\right), \cdots\right\rangle    
\end{equation}
In particular,
\begin{equation}
:\{S_B(z),\overline{S}^A(\bar{z})\}: \mathcal{O}_{\Delta,\ell}(w,\bar{w})= \frac{1}{z-w}\frac{1}{\bar{z}-\bar{w}}\mathcal{O}_{\Delta+1,\ell}(w,\bar{w})+\text{regular.}  
\end{equation}
Comparing this OPE with \eqref{eq:curlyPope} readily implies the desired result 
\begin{equation}
:\{S_B(z),\overline{S}^A(\bar{z})\}:~=\delta^A_B\mathcal{P}(z,\Bar{z}).    
\end{equation}
\subsection{Possible R-symmetry current}\label{sec:R-gencompcalc}
Recall that R-symmetry acts on supercharges $Q^A_{\alpha}$ and $\overline{Q}_{\dot{\alpha}A},~A=1,\dots,8$ by multiplying a unitary matrix $U\in U(8)$. This means that the supercharges transform in the fundamental representation of R-symmetry group. At the level of Lie algebra, we can identify the R-symmetry group as simply $\mathfrak{su}(8)\oplus\mathfrak{u}(1)$ since $\mathrm{U}(8)\cong (\mathrm{SU}(8)\times \mathrm{U}(1))/\mathbb{Z}_8$. Thus we can label the generators of R-symmetry to be $T^A_B$ and $R$, where $T^A_B$ are generators of the fundamental representation of $\mathrm{SU}(8)$ satisfying the $\mathfrak{su}(8)$ algebra:
\begin{equation}
 \left[T^A_B,T^C_D\right]=\delta_{D}^AT^C_B-\delta_{B}^CT^A_D,    
\end{equation}
and $R$ is the generator of the scaling U(1). A suitable matrix representation for the generators is \cite{Bianchi:2008pu},
\begin{equation}
\left(T_{B}^{A}\right)^{C}{ }_{D}=\delta_{D}^{A} \delta_{B}^{C}-\frac{1}{8} \delta_{B}^{A} \delta_{D}^{C}    
\end{equation}
$T^A_B$ acts on the supercharges as 
\begin{equation}
    \left[T^A_B,Q^C_{\alpha}\right]=\left(T^A_B\right)^C_{~D}Q^D_{\alpha},\quad \left[T^A_B,\overline{Q}_{\dot{\alpha}C}\right]=-\left(T^A_B\right)^D_{~C}\overline{Q}_{\dot{\alpha}D}.
\end{equation}
We now want to construct a current $\widetilde{\mathcal{G}}^A_B(z,\bar{z})$ whose modes will extend the generators $T^A_B$. 
As will be shown in Section \ref{sec:asympsym}, the modes of the supercurrents $S_A,\overline{S}^A$ will extend the supercharges. Since the OPE of currents directly translates to the commutator of their modes within radial quantisation, our currents must satisfy the OPE:
\begin{equation}
\begin{split}
& \widetilde{\mathcal{G}}^A_B(z,\bar{z})S_C(w)\sim ((z-w)~ \text{singularity}) \left(T^A_B\right)^D_{~C}S_D(w),\\&\widetilde{\mathcal{G}}^A_B(z,\bar{z})\overline{S}^C(w)\sim -((\bar{z}-\bar{w})~ \text{singularity}) \left(T^A_B\right)^C_{~D}\overline{S}^D(w).
 \end{split}
 \label{eq:curGreqopeSSbar}
\end{equation}
Note that $S_A,\overline{S}^A$ are holomorphic and antiholomorphic currents respectively, this imposes the condition that the singularities in \eqref{eq:curGreqopeSSbar} be holomorphic and antiholomorphic respectively. As will be shown in Section \ref{sec:asympsym}, nonholomorphic (holomorphic) singularity in the OPE of $\mathcal{G}^A_B(z,\bar{z})$ with $S_C(w)$ ($\overline{S}^C(w)$) results in nonsensical algebra. This requirement will be crucial.\par
The only conformal operator we are left with is the graviphoton operator. Moreover, the leading soft graviphoton operator corresponds to $\Delta=0$ as can be inferred from the poles of the beta function in the OPEs of graviphoton operators with other operators which are summarised in Appendix \ref{app:opes}. It is clear that we must consider the order independent graviphoton double soft limit with opposite helicity, that is their normal ordered commutator (since they are bosonic). Since it contains the factor $\delta^{AB}_{CD}$, as can be seen from the collinear limits, this can be manipulated properly to obtain the $\mathrm{SU}(8)$ generators. Here we consider the most general integral transform corresponding to negative and positive helicity soft graviphotons respectively as,
\begin{equation}\label{eq:GGbardef}   
 \begin{split}
&G_{AB}(z, \bar{z}) = \lim_{\Delta\to 0}\frac{\Delta}{ \pi} \int d^{2} z^{\prime} \frac{1}{\left(z-z^{\prime}\right)^{a}}\frac{1}{\left(\bar{z}-\bar{z}^{\prime}\right)^{b}} \mathcal{O}_{AB; \Delta,-1}(z^{\prime}, \bar{z}^{\prime}) \\
&\overline{G}^{CD}(z, \bar{z}) =\lim_{\Delta\to 0}\frac{\Delta}{\pi} \int d^{2} z^{\prime} \frac{1}{\left(\bar{z}-\bar{z}^{\prime}\right)^{a^{\prime}}} \frac{1}{\left(z-z^{\prime}\right)^{b^{\prime}}} \mathcal{O}^{CD}_{\Delta,+1}(z^{\prime}, \bar{z}^{\prime}).
\end{split}
\end{equation}
One can easily see that we can recover the usual shadow transformation \cite{Fan:2020xjj} by taking specific values of $a$ and $b$. The operators $\mathcal{O}_{AB; 0,-1}$ and $\mathcal{O}^{CD}_{0,+1}$ have conformal weights $(-\frac{1}{2},\frac{1}{2})$ and $(\frac{1}{2}, -\frac{1}{2})$ respectively. Hence scaling transformation reveals the conformal weights of the currents $G_{AB}$ and  $\overline{G}^{CD}$ to be $(a-\frac{3}{2},b-\frac{1}{2})$ and $(b^{\prime}-\frac{1}{2},a^{\prime}-\frac{3}{2})$ respectively.\\
Let us start with the OPE of our new currents $G_{AB}$ and  $\overline{G}^{CD}$ with any conformal primary operators.
\begin{equation}
\begin{split}
    &\left\langle G_{AB}(z, \bar{z})\prod_{n=2}^{N}\mathcal{O}^{*_n}_{\Delta_n,\ell_n}(z_n,\bar{z}_n)\right\rangle\\
    &=\lim_{\substack{\Delta_1\to0}}\frac{\Delta_1}{\pi}\int d^2z_1 \frac{1}{\left(z-z_1\right)^{a}}\frac{1}{\left(\bar{z}-\bar{z}_1\right)^{b}} \left\langle\mathcal{O}_{AB~\Delta_1,-1}(z_1,\bar{z}_1)\prod_{n=2}^{N}\mathcal{O}^{*_n}_{\Delta_n,\ell_n}(z_n,\bar{z}_n)\right\rangle
    \\&=\frac{1}{\pi}\int d^2z_1\frac{1}{\left(z-z_1\right)^{a}}\frac{1}{\left(\bar{z}-\bar{z}_1\right)^{b}}\Bigg[\sum_{n=2}^{N}f(A,B,\ell_i,*_n,*_n') \frac{z_1-z_n}{\bar{z}_1-\bar{z}_n}\left\langle \cdots \mathcal{O}^{*'_n}_{\Delta_n,\ell_n +1}(z_n,\bar{z}_n)\right\rangle \Bigg]
\end{split}    
\end{equation}
where $f(A,B,\ell_i,*_n,*_n')$ contains the R-symmetry index of the operators in the correlation function which appears on taking the collinear limit. We used the fact that $\lim_{\Delta\to 0}\Delta \;B(\Delta,*)=1$. Now we use two basic integrals (see \cite[Appendix B]{Jiang:2021ovh} for proof):
\begin{equation}
\begin{split}
&\int d^2 z_1 \frac{1}{\left(z-z_1\right)^A} \frac{1}{\left(\bar{z}-\bar{z}_1\right)^B} \frac{\left(\bar{z}_1-\bar{z}_j\right)^s}{z_1-z_j}=C_s (A, B) \frac{1}{\left(z_j-z\right)^A\left(\bar{z}_j-\bar{z}\right)^{B-s-1}} \\
&\int d^2 z_1 \frac{1}{\left(\bar{z}-\bar{z}_1\right)^A} \frac{1}{\left(z_1-z_1\right)^B} \frac{\left(z_1-z_j\right)^s}{\bar{z}_1-\bar{z}_j}=C_s (A, B) \frac{1}{\left(\bar{z}_j-\bar{z}\right)^A} \frac{1}{\left(z_j-z\right)^{B-s-1}}
\end{split}
\label{eq:genintab}
\end{equation}
where
\begin{equation}\label{st}
C_s (A, B)=\frac{(-1)^{s+A+B}(-\pi) s !}{(-B+1)(-B+2) \cdots(-B+s+1)}\end{equation}
Now performing the shadow integral for $ n\neq1$ and $s=1$,
\begin{equation}\label{shad}
    \begin{split}
        \int d^2z_1\frac{1}{\left(z-z_1\right)^{a}}\frac{1}{\left(\bar{z}-\bar{z}_1\right)^{b}}\frac{z_1-z_n}{\bar{z}_1-\bar{z}_n} =C_1(b,a) \frac{1}{(\bar{z}_n-\bar{z})^b}\frac{1}{(z_n-z)^{a-2}}
    \end{split}
\end{equation}
We have,
\begin{equation}\label{G}
\begin{split}
    &\left\langle G_{AB}(z,\bar{z})\prod_{n=2}^{N}\mathcal{O}^{*_n}_{\Delta_n,\ell_n}(z_n,\bar{z}_n)\right\rangle\\
    &=\sum_{i=2}^{N}f(A,B,\ell_i,*_i,*_i')C_1 (b,a) \frac{1}{(\bar{z}_i-\bar{z})^b}\frac{1}{(z_i-z)^{a-2}} \left\langle \cdots \mathcal{O}^{*'_i}_{\Delta_i,\ell_i +1}(z_i,\bar{z}_i)\right\rangle
\end{split}    
\end{equation}
Here the helicities of the conformal operators inside the correlator are restricted to $\ell_n \in \{ -2, -\frac{3}{2}, -1,-\frac{1}{2}, 0, +\frac{1}{2}, +1\}$. This can be verified from the beta function singularities in the OPEs in Appendix \ref{app:opes}.
Similarly we can have the OPE for antiholomorphic current $\overline{G}^{CD}$ which act on the conformal primaries with helicities restricted in the range $\ell'_n \in \{ -1, -\frac{1}{2},0, +\frac{1}{2}, +1, +\frac{3}{2}, +2\}$,
\begin{equation}\label{barG}
\begin{split}
    &\left\langle \overline{G}^{CD}(z,\bar{z})\prod_{n=2}^{N}\mathcal{O}^{*_n}_{\Delta_n,\ell'_n}(z_n,\bar{z}_n)\right\rangle\\
    &=\sum_{n=2}^{N}\bar{f}(C,D,\ell_i',*_n,*_n')C (b^{\prime},a^{\prime}) \frac{1}{(\bar{z}_n-\bar{z})^{b^{\prime}}}\frac{1}{(z_n-z)^{a^{\prime}-2}}\left\langle \cdots \mathcal{O}^{*'_n}_{\Delta_n,\ell'_n -1}(z_n,\bar{z}_n)\right\rangle
\end{split}    
\end{equation}
Here we can pair $\ell'$ with $\ell=\ell'+1$. Hence we can write the OPEs as,
\begin{equation}
    \begin{split}
        &G_{AB}(z)\mathcal{O}^{*}_{\Delta,\ell}(w,\bar{w}) \sim  f(A,B,\ell,*,*') C_1(b,a) \frac{1}{(\bar{w}-\bar{z})^b}\frac{1}{(w-z)^{a-2}} \mathcal{O}^{*'}_{\Delta,\ell'}(z,\bar{w})\\
        &\overline{G}^{CD}(z)\mathcal{O}^{*}_{\Delta,\ell'}(w,\bar{w}) \sim  \bar{f}(C,D,\ell',*,*') C_1 (b^{\prime},a^{\prime}) \frac{1}{(\bar{w}-\bar{z})^{b^{\prime}}}\frac{1}{(z-z)^{a^{\prime}-2}} \mathcal{O}^{*'}_{\Delta,\ell}(w,\bar{w})
    \end{split}
\end{equation}
\subsubsection{The composite current}
To construct a suitable current for R-symmetry, we need to use double soft limits of the graviphoton operators. As is well known, double soft limit of opposite helicity operators depend on the order of the soft limit. For this reason, as in \cite{Banerjee:2021uxe} we consider the following operator:
\begin{equation}
\mathcal{G}_{AB}^{CD}(z, \bar{z} ; w, \bar{w}):=G_{AB}(z) \overline{G}^{CD}(\bar{w})-\overline{G}^{CD}(\bar{w}) G_{AB}(z) \equiv \Big[G_{AB}(z), \overline{G}^{CD}(\bar{w})\Big].    
\end{equation}
In order to construct a local operator, one needs to consider the normal order of this operator evaluated at $z=w,\Bar{z}=\Bar{z}$. We thus define
\begin{equation}
\begin{split}
    \mathcal{G}_{AB}^{CD}(z,\bar{z})=\;:\mathcal{G}_{AB}^{CD}(z, \bar{z} ; z, \bar{z}):\; &= \; :G_{AB}(z) \overline{G}^{CD}(\bar{z})-\overline{G}^{CD}(\bar{z}) G^{AB}(z):\\& \equiv \;  :\Big[G_{AB}(z), \overline{G}^{CD}(\bar{z})\Big]:.
\end{split}    
\end{equation}
We show in Appendix \ref{intco} that subject to the requirement of the R-symmetry current explained above \eqref{eq:curGreqopeSSbar}, the current $\mathcal{G}_{AB}^{CD}(z,\bar{z})$ satisfies the following OPE:
\begin{equation}\label{compon}
    \begin{split}
&\left\langle \mathcal{G}_{AB}^{CD}(z, \bar{z} )\prod_{n=3}^{N}\mathcal{O}^{*_n}_{\Delta_n,\ell_n}(z_n,\bar{z}_n)\right\rangle\\
&=  (-1)^{a+b+a^{\prime}+b^{\prime}}C_1(b, a)C_1(b^{\prime}, a^{\prime})\Bigg[f(A,B,\ell_j,*_j,*_j')\bar{f}(C,D,\ell_j+1,*_j',*_j'')\\
  &\hspace{1cm}\times\frac{1}{\left(z-z_j\right)^{a+b^{\prime}-2}} \frac{1}{\left(\bar{z}-\bar{z}_j\right)^{a^{\prime}+b-2}}\left\langle\mathcal{O}^{*_3}_{\Delta_3,\ell_3}(z_3,\bar{z}_3) \cdots \mathcal{O}^{*^{''}_j}_{\Delta_j,\ell_j}(z_j,\bar{z}_j)\cdots\mathcal{O}^{*_N}_{\Delta_N,\ell_N}(z_N,\bar{z}_N) \right\rangle\\
  &\hspace{2cm}-\bar{f}(C,D,\ell_j,*_j,*_j')f(A,B,\ell_j-1,*_j',*_j'')\\
  &\hspace{1cm}\times\frac{1}{\left(z-z_j\right)^{a^{\prime}+b-2}} \frac{1}{\left(\bar{z}-\bar{z}_j\right)^{a+b^{\prime}-2}}\left\langle\mathcal{O}^{*_3}_{\Delta_3,\ell_3}(z_3,\bar{z}_3) \cdots \mathcal{O}^{*^{''}_j}_{\Delta_j,\ell_j}(z_j,\bar{z}_j)\cdots\mathcal{O}^{*_N}_{\Delta_N,\ell_N}(z_N,\bar{z}_N) \right\rangle\Bigg]
\end{split}
\end{equation}    
In particular, 
\begin{equation}
\begin{split}
 & \mathcal{G}_{AB}^{CD}(z, \bar{z} )\mathcal{O}_{E~\Delta,-\frac{3}{2}}(w,\bar{w})\sim -\delta^{CD}_{AB} \frac{(-1)^{a+b+a^{\prime}+b^{\prime}}C_1(b, a)C_1(b^{\prime}, a^{\prime})}{\left(z-w\right)^{a+b^{\prime}-2}\left(\bar{z}-\bar{w}\right)^{a^{\prime}+b-2}}  \mathcal{O}_{E~\Delta,-\frac{3}{2}}(w,\bar{w})\\
&\mathcal{G}_{AB}^{CD}(z, \bar{z} )\mathcal{O}^E_{\Delta,+\frac{3}{2}}(w,\bar{w})\sim \delta^{CD}_{AB} \frac{(-1)^{a+b+a^{\prime}+b^{\prime}}C_1(b, a)C_1(b^{\prime}, a^{\prime})}{\left(z-w\right)^{a^{\prime}+b-2}\left(\bar{z}-\bar{w}\right)^{a+b^{\prime}-2}} \mathcal{O}^E_{\Delta,+\frac{3}{2}}(w,\bar{w})
  \end{split}
\end{equation}
where we used the fact that for the gravitino operator, the R-symmetry factor in the double soft limit is $-\delta^{CD}_{AB}$. Indeed
\[
\lim _{\Delta_1 \to 0} \Delta_1 \mathcal{O}_{A B ; \Delta_1,-1}(z, \bar{z}) \mathcal{O}_{E, \Delta,-\frac{3}{2}}\left(z_1, \bar{z}_1\right)=\frac{z-z_1}{\bar{z}-\bar{z}_1} \mathcal{O}_{A B E ; \Delta,-\frac{1}{2}}\left(z_1, \bar{z}_1\right)
\]
If $E=r$, $A=a$, $B=b$, $C=c$, $D=d$, then
\[ 
\lim _{\Delta_2\to 0} \Delta_2 \mathcal{O}_{\Delta_{2,+1}}^{c d}(w, \bar{w}) \mathcal{O}_{a b r ; \Delta,-\frac{1}{2}}\left(z_1, \bar{z}_1\right)=-\delta_{a b}^{c d} \frac{\bar{w}-\bar{z}_1}{w-z_1} \mathcal{O}_{r, \Delta,-\frac{3}{2}}\left(z_1, \bar{z}_1\right)
\]
In all other cases, one can check from the collinear limit in Appendix \ref{app:opes} that the R-symmetry factor is $-\delta^{CD}_{AB}$. 
Let us construct a new current as a linear combination of our previous currents as follows:
\begin{equation}\label{def}
   \left(\widetilde{\mathcal{G}}^{A}_{B}\right)^C_D(z,\bar{z}):=-\Bigg(\frac{1}{7}\delta^A_D\sum_{E=1}^8\mathcal{G}^{EC}_{BE}(z,\bar{z})+\frac{1}{56}\delta_B^A\sum_{E=1}^8\mathcal{G}^{EC}_{ED}(z,\bar{z})\Bigg).
\end{equation}
Using the definition of generalised Kronecker delta
\begin{equation}
    \delta^{a_1\dots a_n}_{b_1\dots b_n}=\sum_{\sigma\in S_n}\text{sign}(\sigma)\delta^{a_1}_{b_{\sigma(1)}}\dots\delta^{a_n}_{b_{\sigma(n)}},
\end{equation}
we see that 
\begin{equation}
   \sum_{E=1}^8\delta^{EC}_{BE}=-7\delta^C_B,\quad \sum_{E=1}^8\delta^{EC}_{ED}=7\delta^C_D. 
\end{equation}
This gives us the OPE
\begin{equation}
\begin{split}
 & \left(\widetilde{\mathcal{G}}^{C}_{A}\right)^D_B(z, \bar{z} )\mathcal{O}_{D~\Delta,-\frac{3}{2}}(w,\bar{w})\sim  \frac{(-1)^{a+b+a^{\prime}+b^{\prime}}C_1(b, a)C_1(b^{\prime}, a^{\prime})}{\left(z-w\right)^{a+b^{\prime}-2}\left(\bar{z}-\bar{w}\right)^{a^{\prime}+b-2}}  \left(T_A^C\right)_B^D\mathcal{O}_{D~\Delta,-\frac{3}{2}}(w,\bar{w})\\
&\left(\widetilde{\mathcal{G}}^{C}_{A}\right)^D_B(z, \bar{z} )\mathcal{O}^B_{\Delta,+\frac{3}{2}}(w,\bar{w})\sim- \frac{(-1)^{a+b+a^{\prime}+b^{\prime}}C_1(b, a)C_1(b^{\prime}, a^{\prime})}{\left(z-w\right)^{a^{\prime}+b-2}\left(\bar{z}-\bar{w}\right)^{a+b^{\prime}-2}} \left(T_A^C\right)_B^D\mathcal{O}^B_{\Delta,+\frac{3}{2}}(w,\bar{w})
  \end{split}
\end{equation}
Hence $\widetilde{\mathcal{G}}^{C}_{A}$ is a candidate which can extend the R-symmetry algebra. But we see an immediate problem. The OPE of $\widetilde{\mathcal{G}}^{C}_{A}$ with supercurrents $S_D(w),\overline{S}^B(\bar{w})$ is given by \begin{equation}
\begin{split}
\left(\widetilde{\mathcal{G}}^{C}_{A}\right)^D_B(z, \bar{z} )S_D(w)\sim & \left(T_A^C\right)_B^D\lim _{\Delta \rightarrow \frac{1}{2}} \frac{\Delta-\frac{1}{2}}{\pi}  (-1)^{a+b+a^{\prime}+b^{\prime}}C_1(b, a)C_1(b^{\prime}, a^{\prime})\\&\times
\int d^2 z_1 \frac{1}{\left(w-z_1\right)^3}
\frac{1}{\left(z-z_1\right)^{a+b^{\prime}-2}} \frac{1}{\left(\bar{z}-\bar{z}_1\right)^{a^{\prime}+b-2}}\mathcal{O}_{D, \Delta,-\frac{3}{2}}\left(z_1, \bar{z}_1\right) 
\end{split}
\label{eq:curlyGS}
\end{equation} 
and 
\begin{equation}
\begin{split}
\left(\widetilde{\mathcal{G}}^{C}_{A}\right)^D_B(z, \bar{z} )\overline{S}^B(\bar{w})\sim & -\left(T_A^C\right)_B^D\lim _{\Delta \rightarrow \frac{1}{2}} \frac{\Delta-\frac{1}{2}}{\pi}  (-1)^{a+b+a^{\prime}+b^{\prime}}C_1(b, a)C_1(b^{\prime}, a^{\prime})\\&\times
\int d^2 z_1 \frac{1}{\left(\bar{w}-\bar{z}_1\right)^3}
\frac{1}{\left(z-z_1\right)^{a'+b-2}} \frac{1}{\left(\bar{z}-\bar{z}_1\right)^{a+b'-2}}\mathcal{O}^B_{ \Delta,+\frac{3}{2}}\left(z_1, \bar{z}_1\right) 
\end{split}  
\label{eq:curlyGSbar}
\end{equation}
The requirement \eqref{eq:curGreqopeSSbar} forces $a'+b-2=0$ in \eqref{eq:curlyGS} and  \eqref{eq:curlyGSbar}. But then in view of \eqref{state} we get 
\begin{equation}\label{demand}
    a+b^{\prime}-2=0 \quad \text{and} \quad a^{\prime}+b-2=0
\end{equation}
and conclude that the OPE is trivial
\begin{equation}
    \left(\widetilde{\mathcal{G}}^{C}_{A}\right)^D_B(z, \bar{z}) S_D(w) \sim \text{regular},\quad \left(\widetilde{\mathcal{G}}^{C}_{A}\right)^D_B(z, \bar{z}) \overline{S}^B(\bar{w}) \sim \text{regular.}
\end{equation}
\section{The $\mathcal{N}=8$  $\mathfrak{sbms}_4$ algebra}\label{sec:asympsym}
Let us now find out the asymptotic symmetries of the theory. The usual symmetry currents in the theory are the stress tensor $T(z),\overline{T}(\bar{z})$, which are the superrotation generators and $\mathcal{P}(z,\bar{z})$, which is the supertranslation generator. The modes of these currents generate the $\mathfrak{bms}_4$ algebra as described in \cite{Fotopoulos:2019vac}. As usual the generators of $\mathfrak{bms}_4$ are the modes of $T(z),\overline{T}(\bar{z})$ and  $\mathcal{P}(z,\bar{z})$. Let us expand these currents in modes:
\begin{equation}
    T(z)=\sum_{n\in\mathbb{Z}}L_nz^{-n-2},\quad \overline{T}(\bar{z})=\sum_{n\in\mathbb{Z}}\bar{L}_n\bar{z}^{-n-2};
\end{equation}
\begin{equation}
    \begin{split}
        \mathcal{P}(z, \bar{z}) \equiv \sum_{n, m \in \mathbb{Z}} P_{n-\frac{1}{2}, m-\frac{1}{2}} z^{-n-1} \bar{z}^{-m-1}
    \end{split}
\end{equation}
As discussed in \cite{Fotopoulos:2019vac}, the modes $P_{n-\frac{1}{2}, m-\frac{1}{2}}$ can be obtained from the modes of the current $P(z)$ or $\overline{P}(\bar{z})$. If we write  
\begin{equation}
    P(z)=\sum_{n\in\mathbb{Z}}P_{n-\frac{1}{2}}z^{-n-1},\quad \overline{P}(\bar{z})=\sum_{m\in\mathbb{Z}}\overline{P}_{m-\frac{1}{2}}z^{-m-1}
\end{equation}
then 
\begin{equation}
    P_{n-\frac{1}{2}, -\frac{1}{2}}=P_{n-\frac{1}{2}},\quad P_{-\frac{1}{2}, m-\frac{1}{2}}=\overline{P}_{m-\frac{1}{2}}
\end{equation}
and 
\begin{equation}
\begin{split}
    P_{n-\frac{1}{2}, m-\frac{1}{2}}&=\frac{1}{i \pi(m+1)} \oint d \bar{w} \bar{w}^{m+1}\left[\bar{T}(\bar{w}), P_{n-\frac{1}{2},-\frac{1}{2}}\right]\\&=\frac{1}{i \pi(m+1)} \oint d w w^{n+1}\left[T(w), \overline{P}_{-\frac{1}{2},m-\frac{1}{2}}\right].
\end{split}
\end{equation}
These modes satisfy the usual $\mathfrak{bms}_4$ algebra:
\begin{equation}\label{bms4}
    \begin{split}
        &\left[L_{m}, L_{n}\right]=(m-n) L_{m+n},\quad \left[\bar{L}_{m}, \bar{L}_{n}\right]=(m-n) \bar{L}_{m+n}\\& \left[L_{n}, P_{k l}\right]=\left(\frac{1}{2} n-k\right) P_{n+k, l},\quad \left[\bar{L}_{n}, P_{k l}\right]=\left(\frac{1}{2} n-l\right) P_{k, n+l},
    \end{split}
\end{equation}
where $m,n\in\mathbb{Z}$ and $k,l\in\mathbb{Z}+\frac{1}{2}$.
In addition, an infinite dimensional extension of the $\mathcal{N}=1$ supersymmetry algebra was constructed in \cite{Fotopoulos:2020bqj}. The supercurrent was shown to be the shadow transform of the gravitino operator. In our theory, we have 8 supercurrents $S^A(z)$ and their antiholomorphic counterpart $\overline{S}_A(\bar{z})$. The OPEs \eqref{eq:TSope} show that $S^A(z)$ and $\overline{S}_A(\bar{z})$ are conformal primaries of dimension $(\frac{3}{2},0)$ and $(0,\frac{3}{2})$ respectively. Consequently if we expand the supercurrents as 
\begin{equation}
\begin{split}
 &S_A(z)=\sum_{k \in \mathbb{Z}+\frac{1}{2}} \frac{(S_A)_k}{z^{k+\frac{3}{2}}}, \quad \text{with} \quad (S_A)_k=\frac{1}{2 \pi i }\oint d z\; z^{k+\frac{1}{2}}\; S_A(z).\\
&\overline{S}^{A}(\bar{z})=\sum_{l \in \mathbb{Z}+\frac{1}{2}} \frac{\overline{S}^A_l}{z^{l+\frac{3}{2}}}, \quad \text{with} \quad \overline{S}^{A}_l=\frac{1}{2 \pi i }\oint d \bar{z}\; \bar{z}^{l+\frac{1}{2}} \;\overline{S}^{A}(\bar{z})
    \end{split}
\end{equation}
then we can write the commutator of these modes with the Virasoro generators as:
\begin{equation}\label{ls}
\begin{split}
    &\left[L_n,(S_A)_{m}\right]=\left(\frac{n}{2}-m\right)(S_A)_{m+n},\quad \left[\bar{L}_n,(S_A)_{m}\right]=0\\&\left[L_n,\overline{S}^A_{m}\right]=0,\quad \left[\bar{L}_n,\overline{S}^A_{m}\right]=\left(\frac{n}{2}-m\right)\overline{S}^A_{m+n}.
    \end{split}
\end{equation}
The operator relation \eqref{eq:SSbarnormord} gives the anticommutator
\begin{equation}
   \{(S_B)_m,\overline{S}^A_n\}=\delta^A_BP_{mn},\quad m,n\in\mathbb{Z}+\frac{1}{2}. 
\end{equation}
Let us now discuss the requirement of (anti)holomorphicity of the singularity in \eqref{eq:curGreqopeSSbar}. Suppose $(\widetilde{\mathcal{G}}^{A}_{B})^C_D(z,\bar{z})$ has conformal weights\footnote{the scaling dimensions of $(\widetilde{\mathcal{G}}^{A}_{B})^C_D(z,\bar{z})$ can be calculated from those of $G_{AB},\overline{G}^{CD}$. It is $(a+b'-2,a'+b-2)$.}  $(h,\bar{h})$. Then we can expand the current as  
\begin{equation}
    (\widetilde{\mathcal{G}}^{A}_{B})^C_D(z,\bar{z})=\sum_{n,m\in\mathbb{Z}}\big{\{}(\widetilde{\mathcal{G}}^{A}_{B})^C_D\big{\}}_{mn }z^{-m-h}\bar{z}^{-n-\bar{h}},
    \label{eq:curlyGmode}
\end{equation}
with 
\begin{equation}
    \big{\{}(\widetilde{\mathcal{G}}^{A}_{B})^C_D\big{\}}_{m n}=\frac{1}{(2\pi i)^2}\oint dz\oint d\bar{z} ~z^{m+h-1}\bar{z}^{n+\bar{h}+1}(\widetilde{\mathcal{G}}^{A}_{B})^C_D(z,\bar{z}).
\end{equation}
Suppose we had an OPE of the form\footnote{exactly same argument works if we have higher power singularities.} 
\begin{equation}
(\widetilde{\mathcal{G}}^{A}_{B})^C_D(z,\bar{z})S_C(w)\sim \frac{1}{z-w}\frac{1}{\bar{z}-\bar{w}} (T^A_B)^C_DS_C(w).   
\end{equation}
One can readily check that this would give us the commutator
\begin{equation}
\left[(\widetilde{\mathcal{G}}^{A}_{B})^C_D\big{\}}_{m n},(S_C)_k\right]=\bar{w}^n \left(T_B^A\right)_D^C \left\{S_C\right\}_{m+k}    
\end{equation}
This is nonsensical since we do not have any $\bar{w}$ dependence on the left hand side. Similarly one can justify the second part of \eqref{eq:curGreqopeSSbar}. So we conclude that the $\mathcal{N}=8~\mathfrak{sbms}_4$ algebra does not contain the extension of global R-symmetry algebra. The final algebra is then given by 
\begin{equation}\label{LG}
\begin{split}
        &\left[L_{m}, L_{n}\right]=(m-n) L_{m+n},\quad \left[\bar{L}_{m}, \bar{L}_{n}\right]=(m-n) \bar{L}_{m+n}\\& \left[L_{n}, P_{k l}\right]=\left(\frac{1}{2} n-k\right) P_{n+k, l},\quad \left[\bar{L}_{n}, P_{k l}\right]=\left(\frac{1}{2} n-l\right) P_{k, n+l},\\&\left[L_n,(S_A)_{m}\right]=\left(\frac{n}{2}-m\right)(S_A)_{m+n},\quad \left[\bar{L}_n,(S_A)_{m}\right]=0\\&\left[L_n,\overline{S}^A_{m}\right]=0,\quad \left[\bar{L}_n,\overline{S}^A_{m}\right]=\left(\frac{n}{2}-m\right)\overline{S}^A_{m+n}\\&\{(S_B)_m,\overline{S}^A_n\}=\delta^A_BP_{mn}. 
    \end{split}    
\end{equation}
\section{Conclusion}\label{sec:conc}
In this paper, we have used the CCFT technique to compute the asymptotic symmetry algebra of $\mathcal{N}=8$ supergravity in asymptotically flat spacetime. The crucial part of our result is the non-extension of the global $\mathrm{SU}(8)_R$ R-symmetry algebra. The purely mathematical considerations \cite{Banerjee:2022abf} for $\mathcal{N}=2$ theory suggests that the infinite dimensional extension of R-symmetry is fraught with mathematical inconsistencies. Here, performing a direct asymptotic symmetry analysis of the supergravity theory using CCFT prescription, we have confirmed that indeed supergravity does not result in such an extension. The rest of the symmetry algebra is as expected and is presented in \eqref{LG}.\par
Let us end the paper with relevant open problems. In the seminal work of Hawking et.al. \cite{Hawking:2016sgy} the importance of the infinite number of soft hairs in the context of black hole microscopics was discussed. The study was further taken forward in \cite{Haco:2018ske, Hawking:2016msc, Hawking:2016sgy, Chu:2018tzu, Averin:2016hhm} and beautifully reviewed in \cite{Strominger:2017zoo}. They emphasized the importance of symmetry enhancements at the future horizon $\mathcal{H}^+$ of the black holes and how both the hypersurfaces\footnote{$\mathcal{I}^+$ denotes the future null horizon} $\mathcal{H}^+$ and $\mathcal{I}^+$ carry information of conserved charges, that are in turn important for understanding black hole microscopics.  The study of the present paper indicates that the asymptotic soft hairs of the supergravity theories will not have distinct infinite R-charges, rather they will only carry the global fixed number of R-charges. An interesting question that remains to be studied is the effect, if any, of these R-charges at the horizon and finally their importance in the black hole microscopics. We hope to return to this question in future.
\section*{Acknowledgement} The authors would like to thank Shamik Banerjee, Arpita Mitra, Debanshu Mukherjee and Ashoke Sen for useful discussions. NB and TR are thankful to participants to Regional String Meeting-Bhopal chapter for sharing interesting outlook relevant for the present study. RKS would like to thank Anindya Banerjee for various useful discussions. The work of RKS is supported by the US Department of Energy under grant DE-SC0010008. Finally we are thankful to the people of India for their generous support to the study of basic sciences.
\begin{appendix}
\section{OPEs of component fields}\label{app:opes}
Using \eqref{eq:betaintgenrel}, we can extract rest of the OPEs from the from the collinear singularities of the amplitudes calculated in \cite{tab}. In the following, the zero, one, two, three and four index operators are respectively graviton, gravitino, graviphoton, graviphotino and scalar operators.
\subsection{Same spin OPEs}
\begin{equation}\label{fermion}
\begin{split}
&\mathcal{O}^{A}_{\Delta_{1},+\frac{3}{2}}\left(z_{1}, \bar{z}_{1}\right) \mathcal{O}^{B}_{\Delta_{2},+\frac{3}{2}}\left(z_{2,} \bar{z}_{2}\right) \sim \frac{\bar{z}_{12}}{z_{12}} B\left(\Delta_{1}-\frac{1}{2}, \Delta_{2}-\frac{1}{2}\right) \mathcal{O}^{AB}_{\Delta_{1}+\Delta_{2},+1}\left(z_{2}, \bar{z}_{2}\right)\\
&\mathcal{O}^{A}_{\Delta_{1},+\frac{3}{2}}\left(z_{1}, \bar{z}_{1}\right) \mathcal{O}_{B \; \Delta_{2},-\frac{3}{2}}\left(z_{2,} \bar{z}_{2}\right) \sim \frac{\bar{z}_{12}}{z_{12}} \; \delta^{A}_{B} \;B\left(\Delta_{1}-\frac{1}{2}, \Delta_{2}+\frac{5}{2}\right) \mathcal{O}_{\Delta_{1}+\Delta_{2},-2}\left(z_{2}, \bar{z}_{2}\right)\\
&\hspace{35ex} +\frac{z_{12}}{\bar{z}_{12}}\; \delta^{A}_{B} \; B\left(\Delta_{1}+\frac{5}{2}, \Delta_{2}- \frac{1}{2}\right) \mathcal{O}_{\Delta_{1}+\Delta_{2},+2}\left(z_{2}, \bar{z}_{2}\right)
\end{split}
\end{equation}
\begin{equation}
    \begin{split}
        &\mathcal{O}^{AB}_{\Delta_{1},+1}\left(z_{1}, \bar{z}_{1}\right) \mathcal{O}^{CD}_{\Delta_{2},+1}\left(z_{2,} \bar{z}_{2}\right) \sim \frac{\bar{z}_{12}}{z_{12}} B\left(\Delta_{1}, \Delta_{2}\right) \mathcal{O}^{ABCD}_{\Delta_{1}+\Delta_{2},0}\left(z_{2}, \bar{z}_{2}\right)\\
        &\mathcal{O}^{AB}_{\Delta_{1},+1}\left(z_{1}, \bar{z}_{1}\right) \mathcal{O}_{CD; \; \Delta_{2},-1}\left(z_{2,} \bar{z}_{2}\right) \sim -\delta^{AB}_{CD}\Big[ \frac{\bar{z}_{12}}{z_{12}} B\left(\Delta_{1}, \Delta_{2}+2\right) \mathcal{O}_{\Delta_{1}+\Delta_{2},-2}\left(z_{2}, \bar{z}_{2}\right)\\
        &\hspace{35ex} +\frac{z_{12}}{\bar{z}_{12}} B\left(\Delta_{1}+2, \Delta_{2}\right) \mathcal{O}_{\Delta_{1}+\Delta_{2},+2}\left(z_{2}, \bar{z}_{2}\right)\Big]
    \end{split}
    \label{eq:ope+1-1}
\end{equation}
In the following, the notation is $a,b,c,\dots\in\{1,2,3,4\}$ and $r,s,t,\dots\in \{5,6,7,8\}$. See \cite{tab} for details.
\begin{equation}
    \begin{split}
    &\mathcal{O}^{ars}_{\Delta_{1},+\frac{1}{2}}\left(z_{1}, \bar{z}_{1}\right) \mathcal{O}^{btu}_{\Delta_{2},+\frac{1}{2}}\left(z_{2,} \bar{z}_{2}\right) \sim \epsilon^{rstu} \epsilon^{abcd} \frac{z_{12}}{\bar{z}_{12}} B\left(\Delta_{1}+\frac{1}{2}, \Delta_{2}+\frac{1}{2}\right) \mathcal{O}_{cd;\Delta_{1}+\Delta_{2},-1}\left(z_{2}, \bar{z}_{2}\right)\\
&\mathcal{O}^{ars}_{\Delta_{1},+\frac{1}{2}}\left(z_{1}, \bar{z}_{1}\right) \mathcal{O}^{bct}_{\Delta_{2},+\frac{1}{2}}\left(z_{2,} \bar{z}_{2}\right) \sim \epsilon^{rstu} \epsilon^{abcd} \frac{z_{12}}{\bar{z}_{12}} B\left(\Delta_{1}+\frac{1}{2}, \Delta_{2}+\frac{1}{2}\right) \mathcal{O}_{ud; \Delta_{1}+\Delta_{2},-1}\left(z_{2}, \bar{z}_{2}\right)\\
&\mathcal{O}^{rst}_{\Delta_{1},+\frac{1}{2}}\left(z_{1}, \bar{z}_{1}\right) \mathcal{O}^{abc}_{\Delta_{2},+\frac{1}{2}}\left(z_{2,} \bar{z}_{2}\right) \sim \epsilon^{rstu} \epsilon^{abcd} \frac{z_{12}}{\bar{z}_{12}} B\left(\Delta_{1}+\frac{1}{2}, \Delta_{2}+\frac{1}{2}\right) \mathcal{O}_{ud; \Delta_{1}+\Delta_{2},-1}\left(z_{2}, \bar{z}_{2}\right)\\
&\mathcal{O}^{ars}_{\Delta_{1},+\frac{1}{2}}\left(z_{1}, \bar{z}_{1}\right) \mathcal{O}_{btu;\Delta_{2},-\frac{1}{2}}\left(z_{2,} \bar{z}_{2}\right) \sim \epsilon_{tuvw}\epsilon^{rsvw} \delta^{a}_{b} \Big[ \frac{z_{12}}{\bar{z}_{12}} B\left(\Delta_{1}+\frac{3}{2}, \Delta_{2}+\frac{1}{2}\right) \mathcal{O}_{\Delta_{1}+\Delta_{2},+2}\left(z_{2}, \bar{z}_{2}\right)\\
& \hspace{35ex} + \frac{\bar{z}_{12}}{z_{12}} B\left(\Delta_{1}+\frac{1}{2}, \Delta_{2}+\frac{3}{2}\right) \mathcal{O}_{\Delta_{1}+\Delta_{2},-2}\left(z_{2}, \bar{z}_{2}\right) \Big]
\end{split}
\end{equation}
\begin{equation}
    \begin{split}
&\mathcal{O}^{abrs}_{\Delta_{1},0}\left(z_{1}, \bar{z}_{1}\right) \mathcal{O}^{cdtu}_{\Delta_{2},0}\left(z_{2,} \bar{z}_{2}\right) \sim \epsilon^{abcd}\epsilon^{rstu} B\left(\Delta_{1}+1, \Delta_{2}+1\right) \Big[ \frac{z_{12}}{\bar{z}_{12}}  \mathcal{O}_{\Delta_{1}+\Delta_{2},+2}\left(z_{2}, \bar{z}_{2}\right)\\
& \hspace{40ex} + \frac{\bar{z}_{12}}{z_{12}} \mathcal{O}_{\Delta_{1}+\Delta_{2},-2}\left(z_{2}, \bar{z}_{2}\right) \Big]
\end{split}
\end{equation}
\subsection{Different spins}
\begin{equation}
    \begin{split}
&\mathcal{O}_{\Delta_{1},+2}\left(z_{1}, \bar{z}_{1}\right) \mathcal{O}^{A}_{\Delta_{2},+\frac{3}{2}}\left(z_{2,} \bar{z}_{2}\right) \sim \frac{\bar{z}_{12}}{z_{12}} B\left(\Delta_{1}-1, \Delta_{2}-\frac{1}{2}\right) \mathcal{O}^{A}_{\Delta_{1}+\Delta_{2},+\frac{3}{2}}\left(z_{2}, \bar{z}_{2}\right)\\
&\mathcal{O}_{\Delta_{1},+2}\left(z_{1}, \bar{z}_{1}\right) \mathcal{O}_{A; \Delta_{2},-\frac{3}{2}}\left(z_{2,} \bar{z}_{2}\right) \sim \frac{\bar{z}_{12}}{z_{12}} B\left(\Delta_{1}-1, \Delta_{2}+\frac{5}{2}\right) \mathcal{O}_{A;\Delta_{1}+\Delta_{2},-\frac{3}{2}}\left(z_{2}, \bar{z}_{2}\right)
\end{split}
\end{equation}
\begin{equation}
    \begin{split}
    &\mathcal{O}_{\Delta_{1},+2}\left(z_{1}, \bar{z}_{1}\right) \mathcal{O}^{AB}_{\Delta_{2},+1}\left(z_{2,} \bar{z}_{2}\right) \sim \frac{\bar{z}_{12}}{z_{12}} B\left(\Delta_{1}-1, \Delta_{2}\right) \mathcal{O}^{AB}_{\Delta_{1}+\Delta_{2},+1}\left(z_{2}, \bar{z}_{2}\right)\\
&\mathcal{O}_{\Delta_{1},+2}\left(z_{1}, \bar{z}_{1}\right) \mathcal{O}_{AB; \Delta_{2},-1}\left(z_{2,} \bar{z}_{2}\right) \sim \frac{\bar{z}_{12}}{z_{12}} B\left(\Delta_{1}-1, \Delta_{2}+2\right) \mathcal{O}_{AB;\; \Delta_{1}+\Delta_{2},-1}\left(z_{2}, \bar{z}_{2}\right)
\end{split}
\end{equation}
\begin{equation}
    \begin{split}
&\mathcal{O}_{\Delta_{1},+2}\left(z_{1}, \bar{z}_{1}\right) \mathcal{O}^{abr}_{ \Delta_{2},+\frac{1}{2}}\left(z_{2,} \bar{z}_{2}\right) \sim \frac{\bar{z}_{12}}{z_{12}} B\left(\Delta_{1}-1, \Delta_{2}+\frac{1}{2}\right) \mathcal{O}^{abr}_{\Delta_{1}+\Delta_{2},+\frac{1}{2}}\left(z_{2}, \bar{z}_{2}\right)\\
&\mathcal{O}_{\Delta_{1},+2}\left(z_{1}, \bar{z}_{1}\right) \mathcal{O}^{abc}_{ \Delta_{2},+\frac{1}{2}}\left(z_{2,} \bar{z}_{2}\right) \sim  \frac{\bar{z}_{12}}{z_{12}} B\left(\Delta_{1}-1, \Delta_{2}+\frac{1}{2}\right) \mathcal{O}^{abc}_{\Delta_{1}+\Delta_{2},+\frac{1}{2}}\left(z_{2}, \bar{z}_{2}\right)\\
&\mathcal{O}_{\Delta_{1},+2}\left(z_{1}, \bar{z}_{1}\right) \mathcal{O}_{abc; \Delta_{2},-\frac{1}{2}}\left(z_{2,} \bar{z}_{2}\right) \sim -\frac{\bar{z}_{12}}{z_{12}} B\left(\Delta_{1}-1, \Delta_{2}+\frac{3}{2}\right) \mathcal{O}_{abc; \;\Delta_{1}+\Delta_{2},-\frac{1}{2}}\left(z_{2}, \bar{z}_{2}\right)
\end{split}
\end{equation}
\begin{equation}
    \begin{split}
        \mathcal{O}_{\Delta_{1},+2}\left(z_{1}, \bar{z}_{1}\right) \mathcal{O}^{ABCD}_{ \Delta_{2},0}\left(z_{2,} \bar{z}_{2}\right) \sim \frac{\bar{z}_{12}}{z_{12}} B\left(\Delta_{1}-1, \Delta_{2}+1\right) \mathcal{O}^{ABCD}_{ \;\Delta_{1}+\Delta_{2},0}\left(z_{2}, \bar{z}_{2}\right)
    \end{split}
\end{equation}
\begin{equation}
    \begin{split}
&\mathcal{O}^{A}_{\Delta_{1},+\frac{3}{2}}\left(z_{1}, \bar{z}_{1}\right) \mathcal{O}^{BC}_{ \Delta_{2},+1}\left(z_{2,} \bar{z}_{2}\right) \sim \frac{\bar{z}_{12}}{z_{12}} B\left(\Delta_{1}-\frac{1}{2}, \Delta_{2}\right) \mathcal{O}^{ABC}_{ \;\Delta_{1}+\Delta_{2},+\frac{1}{2}}\left(z_{2}, \bar{z}_{2}\right)\\
&\mathcal{O}^{A}_{\Delta_{1},+\frac{3}{2}}\left(z_{1}, \bar{z}_{1}\right) \mathcal{O}_{BC;\; \Delta_{2},-1}\left(z_{2,} \bar{z}_{2}\right) \sim 2!\; \delta^{A}_{[B}\frac{\bar{z}_{12}}{z_{12}} B\left(\Delta_{1}-\frac{1}{2}, \Delta_{2}\right) \mathcal{O}_{C]; \;\Delta_{1}+\Delta_{2},-\frac{3}{2}}\left(z_{2}, \bar{z}_{2}\right)
\end{split}
\end{equation}
\begin{equation}
    \begin{split}
&\mathcal{O}^{A}_{\Delta_{1},+\frac{3}{2}}\left(z_{1}, \bar{z}_{1}\right) \mathcal{O}^{BCD}_{ \Delta_{2},+\frac{1}{2}}\left(z_{2,} \bar{z}_{2}\right) \sim \frac{\bar{z}_{12}}{z_{12}} B\left(\Delta_{1}-\frac{1}{2}, \Delta_{2}+\frac{1}{2}\right) \mathcal{O}^{ABCD}_{ \;\Delta_{1}+\Delta_{2},0}\left(z_{2}, \bar{z}_{2}\right)\\
&\mathcal{O}^{A}_{\Delta_{1},+\frac{3}{2}}\left(z_{1}, \bar{z}_{1}\right) \mathcal{O}_{BCD \Delta_{2},-\frac{1}{2}}\left(z_{2,} \bar{z}_{2}\right) \sim 3 \frac{\bar{z}_{12}}{z_{12}} B\left(\Delta_{1}-\frac{1}{2}, \Delta_{2}+\frac{3}{2}\right) \delta^A_{[B}\mathcal{O}_{CD]; \;\Delta_{1}+\Delta_{2},-1}\left(z_{2}, \bar{z}_{2}\right)
    \end{split}
\end{equation}
\begin{equation}
    \begin{split}
        &\mathcal{O}^{A}_{\Delta_{1},+\frac{3}{2}}\left(z_{1}, \bar{z}_{1}\right) \mathcal{O}^{BCDE}_{\Delta_{2},0} \left(z_{2}, \bar{z}_{2}\right) \sim -\frac{1}{6}\epsilon^{ABCDEFGH} \frac{\bar{z}_{12}}{z_{12}} B\left(\Delta_{1}-\frac{1}{2}, \Delta_{2}+1\right) \mathcal{O}_{FGH; \;\Delta_{1}+\Delta_{2},-\frac{1}{2}}\left(z_{2}, \bar{z}_{2}\right)\\
        &\mathcal{O}^{A}_{\Delta_{1},+\frac{3}{2}}\left(z_{1}, \bar{z}_{1}\right) \mathcal{O}_{BCDE\Delta_{2},0}\left(z_{2},\bar{z}_{2}\right) \sim 3!\delta^A_{[B} \frac{\bar{z}_{12}}{z_{12}} B\left(\Delta_{1}-\frac{1}{2}, \Delta_{2}+1\right) \mathcal{O}_{CDE]; \;\Delta_{1}+\Delta_{2},-\frac{1}{2}}\left(z_{2}, \bar{z}_{2}\right)
    \end{split}
\end{equation}
    \begin{equation}
    \begin{split}
        &\mathcal{O}^{ab}_{\Delta_{1},+1}\left(z_{1}, \bar{z}_{1}\right) \mathcal{O}^{cdr}_{ \Delta_{2},+\frac{1}{2}}\left(z_{2,} \bar{z}_{2}\right) \sim \frac{1}{3!} \epsilon^{rstu}\epsilon^{abcd} \frac{\bar{z}_{12}}{z_{12}} B\left(\Delta_{1}, \Delta_{2}+\frac{1}{2}\right) \mathcal{O}_{stu; \;\Delta_{1}+\Delta_{2},-\frac{1}{2}}\left(z_{2}, \bar{z}_{2}\right)\\
        &\mathcal{O}^{AB}_{\Delta_{1},+1}\left(z_{1}, \bar{z}_{1}\right) \mathcal{O}_{CDE; \; \Delta_{2},-\frac{1}{2}}\left(z_{2,} \bar{z}_{2}\right) \sim -\delta^{AB}_{CD}  \frac{\bar{z}_{12}}{z_{12}} B\left(\Delta_{1}, \Delta_{2}+\frac{3}{2}\right) \mathcal{O}_{E; \;\Delta_{1}+\Delta_{2},-\frac{3}{2}}\left(z_{2}, \bar{z}_{2}\right)
    \end{split}
\end{equation}
 \begin{equation}
    \begin{split}
        &\mathcal{O}^{ab}_{\Delta_{1},+1}\left(z_{1}, \bar{z}_{1}\right) \mathcal{O}^{cdrs}_{ \Delta_{2},0}\left(z_{2,} \bar{z}_{2}\right) \sim \epsilon^{rstu}\epsilon^{abcd} \frac{\bar{z}_{12}}{z_{12}} B\left(\Delta_{1}, \Delta_{2}+1\right) \mathcal{O}_{tu; \;\Delta_{1}+\Delta_{2},-1}\left(z_{2}, \bar{z}_{2}\right)\\
        &\mathcal{O}^{ab}_{\Delta_{1},+1}\left(z_{1}, \bar{z}_{1}\right) \mathcal{O}^{cdef}_{ \Delta_{2},0}\left(z_{2,} \bar{z}_{2}\right) \sim -\epsilon^{cdef}\epsilon^{abgh} \frac{\bar{z}_{12}}{z_{12}} B\left(\Delta_{1}, \Delta_{2}+1\right) \mathcal{O}_{gh; \;\Delta_{1}+\Delta_{2},-1}\left(z_{2}, \bar{z}_{2}\right)
    \end{split}
\end{equation}
   \begin{equation}
    \begin{split}
        &\mathcal{O}^{abr}_{\Delta_{1},+\frac{1}{2}}\left(z_{1}, \bar{z}_{1}\right) \mathcal{O}^{cdst}_{ \Delta_{2},0}\left(z_{2,} \bar{z}_{2}\right) \sim \epsilon^{rstu}\epsilon^{abcd} \frac{\bar{z}_{12}}{z_{12}} B\left(\Delta_{1}+\frac{1}{2}, \Delta_{2}+1\right) \mathcal{O}_{u; \;\Delta_{1}+\Delta_{2},-\frac{3}{2}}\left(z_{2}, \bar{z}_{2}\right)\\
        &\mathcal{O}^{abr}_{\Delta_{1},+\frac{1}{2}}\left(z_{1}, \bar{z}_{1}\right) \mathcal{O}^{cstu}_{ \Delta_{2},0}\left(z_{2,} \bar{z}_{2}\right) \sim -\epsilon^{rstu}\epsilon^{abcd} \frac{\bar{z}_{12}}{z_{12}} B\left(\Delta_{1}+\frac{1}{2}, \Delta_{2}+1\right) \mathcal{O}_{d; \;\Delta_{1}+\Delta_{2},-\frac{3}{2}}\left(z_{2}, \bar{z}_{2}\right)
    \end{split}
\end{equation}
Similarly all other OPEs can be extracted from the amplitudes given in \cite{tab}.
\section{OPE of the composite current $\mathcal{G}^{CD}_{AB}(z,\Bar{z})$}\label{intco}
We begin by calculating the OPEs $G\overline{G}$. We have\footnote{Since there is also an overlap as there in case of soft gravitino currents in section \ref{sec:SbarS} we do not seperate the operators in the correlator according to their spins $\ell,\ell^{'}$ and keep the spins to be arbitrary here as well.}
\begin{equation}
\begin{split}
&\left\langle G_{AB}(z, \bar{z})\overline{G}^{CD}(w,\bar{w})\prod_{n=3}^N\mathcal{O}^{*_n}_{\Delta_n,\ell_n}(z_n,\bar{z}_n)\right\rangle\\&=\lim_{\substack{\Delta_1\to 0\\\Delta_2\to 0}}\frac{\Delta_1\Delta_2}{\pi^2}\int d^2z_1\frac{1}{\left(z-z_1\right)^a} \frac{1}{\left(\bar{z}-\bar{z}_1\right)^b} \frac{1}{\left(\bar{w}-\bar{z}_2\right)^{a^{\prime}}} \frac{1}{\left(w-z_2\right)^{b^{\prime}}}\\
&\hspace{2cm}\times \left\langle\mathcal{O}_{AB~\Delta_1,-1}(z_1,\bar{z}_1)\mathcal{O}^{CD}_{\Delta_2,+1}(z_2,\bar{z}_2)\prod_{n=3}^N\mathcal{O}^{*_n}_{\Delta_n,\ell_n}(z_n,\bar{z}_n)\right\rangle
    \end{split}
\end{equation}
 By taking the soft limit of the first operator $\Delta_1\to 0$,
\begin{equation}
\begin{split}
&\left\langle G_{AB}(z)\overline{G}^{CD}(\bar{w})\prod_{n=3}^N\mathcal{O}^{*_n}_{\Delta_n,\ell_n}(z_n,\bar{z}_n)\right\rangle\\&=\lim_{\substack{\Delta_2\to 0}}\frac{ \Delta_2}{\pi^2}\int d^2z_1\int d^2z_2\frac{1}{\left(z-z_1\right)^a} \frac{1}{\left(\bar{z}-\bar{z}_1\right)^b} \frac{1}{\left(\bar{w}-\bar{z}_2\right)^{a^{\prime}}} \frac{1}{\left(w-z_2\right)^{b^{\prime}}}\\
    &\times \Bigg[ -\delta^{CD}_{AB}\frac{z_1-z_2}{\bar{z}_1-\bar{z}_2}\left\langle\mathcal{O}_{\Delta_2,+2}(z_2,\bar{z}_2)\prod_{n=3}^N\mathcal{O}^{*_n}_{\Delta_n,\ell_n}(z_n,\bar{z}_n)\right\rangle \\&+ \sum_{j=3}^{N}f(A,B,\ell_j,*_j,*_j') \frac{z_1-z_j}{\bar{z}_1-\bar{z}_j}\left\langle\mathcal{O}^{CD}_{\Delta_2,+1}(z_2,\bar{z}_2) \cdots \mathcal{O}^{*'_j}_{\Delta_j,\ell_j+1}(z_j,\bar{z}_j)\cdots\mathcal{O}^{*_N}_{\Delta_N,\ell_N}(z_N,\bar{z}_N) \right\rangle \Bigg]
\end{split}    
\end{equation}\label{oppo}
Now doing the first integral using \eqref{eq:genintab}, we get
\begin{equation}\label{comresu}
\begin{split}
    &\left\langle G_{AB}(z)\overline{G}^{CD}(\bar{w})\prod_{n=3}^N\mathcal{O}^{*_n}_{\Delta_n,\ell_n}(z_n,\bar{z}_n)\right\rangle\\&=\lim_{\substack{\Delta_2\to 0}}\frac{\Delta_2}{\pi }\int d^2z_2 \frac{1}{\left(\bar{w}-\bar{z}_2\right)^{a^{\prime}}} \frac{1}{\left(w-z_2\right)^{b^{\prime}}}\\
    &\times \Bigg[ -\delta^{CD}_{AB}C_1(b, a) \frac{1}{\left(z_2-z\right)^{a-2}\left(\bar{z}_2-\bar{z}\right)^b}\left\langle\mathcal{O}_{\Delta_2,+2}(z_2,\bar{z}_2)\prod_{n=3}^{N}\mathcal{O}^{*_n}_{\Delta_n,\ell_n}(z_n,\bar{z}_n)\right\rangle \\&+ \sum_{j=3}^{N}f(A,B,\ell_j,*_j,*_j') C_1(b, a) \frac{1}{\left(z_j-z\right)^{a-2}} \frac{1}{\left(\bar{z}_j-\bar{z}\right)^b}\\
    &\hspace{5cm}\times\left\langle\mathcal{O}^{CD}_{\Delta_2,+1}(z_2,\bar{z}_2) \cdots \mathcal{O}^{*'_j}_{\Delta_j,\ell_j+1}(z_j,\bar{z}_j)\cdots\mathcal{O}^{*_N}_{\Delta_N,\ell_N}(z_N,\bar{z}_N) \right\rangle \Bigg]
\end{split}    
\end{equation}
We now use the collinear limits of the graviton operator with other fields in the first term and take the conformally soft limit $\Delta_2 \to 0$ in the second term. \\
The first term becomes
\begin{equation}
\begin{split}
    &\lim_{\substack{\Delta_2\to 0}}\frac{\Delta_2}{\pi }\int d^2z_2 \frac{1}{\left(\bar{w}-\bar{z}_2\right)^{a^{\prime}}} \frac{1}{\left(w-z_2\right)^{b^{\prime}}}\Bigg[ -\delta^{CD}_{AB}C_1(b, a) \frac{1}{\left(z_2-z\right)^{a-2}\left(\bar{z}_2-\bar{z}\right)^b}\\
    &\hspace{3cm}\times\left\langle\mathcal{O}_{\Delta_2,+2}(z_2,\bar{z}_2)\prod_{n=3}^{N}\mathcal{O}^{*_n}_{\Delta_n,\ell_n}(z_n,\bar{z}_n)\right\rangle \Bigg]\\
    &=-\delta^{CD}_{AB}C_1(b, a)\lim_{\substack{\Delta_2\to 0}}\frac{\Delta_2}{ \pi }\int d^2z_2 \frac{1}{\left(\bar{w}-\bar{z}_2\right)^{a^{\prime}}} \frac{1}{\left(w-z_2\right)^{b^{\prime}}} \frac{1}{\left(z_2-z\right)^{a-2}\left(\bar{z}_2-\bar{z}\right)^b}\\
    &\times\sum_{i=3}^{N}B(\Delta_2-1, f(\Delta_i))\frac{\bar{z}_2- \bar{z}_i}{z_2-z_i}\left\langle\mathcal{O}^{*_3}_{\Delta_3,\ell_3}(z_3,\bar{z}_3)\cdots\mathcal{O}^{*_i}_{\Delta_i,\ell_i}(z_i,\bar{z}_i)\cdots\mathcal{O}^{*_N}_{\Delta_N,\ell_N}(z_N,\bar{z}_N)\right\rangle \\
    &=-\delta^{CD}_{AB}C_1(b, a)\frac{1}{ \pi }\int d^2z_2 \frac{1}{\left(\bar{w}-\bar{z}_2\right)^{a^{\prime}}} \frac{1}{\left(w-z_2\right)^{b^{\prime}}} \frac{1}{\left(z_2-z\right)^{a-2}\left(\bar{z}_2-\bar{z}\right)^b}\\
    &\times \sum_{i=3}^{N}(1-f(\Delta_i))\frac{\bar{z}_2- \bar{z}_i}{z_2-z_i}\times\left\langle\mathcal{O}^{*_3}_{\Delta_3,\ell_3}(z_3,\bar{z}_3)\cdots\mathcal{O}^{*_i}_{\Delta_i,\ell_i}(z_i,\bar{z}_i)\cdots\mathcal{O}^{*_N}_{\Delta_N,\ell_N}(z_N,\bar{z}_N)\right\rangle 
\end{split}
\end{equation}
where we used 
\[  \begin{split}
 \lim_{\substack{\Delta_2\to 0}}   \Delta_2 \; B(\Delta_2-1, f(\Delta_i))=1-f(\Delta_i) 
\end{split}
\]
In the second term in  \eqref{comresu} now we can take $\Delta_2 \to 0$ limit,
\begin{equation}\label{2nd}
\begin{split}
  &\lim_{\substack{\Delta_2\to 0}}\frac{\Delta_2}{\pi } \sum_{j=3}^{N}f(A,B,\ell_j,*_j,*_j')\frac{1}{\left(z_j-z\right)^{a-2}} \frac{1}{\left(\bar{z}_j-\bar{z}\right)^b}\int d^2z_2 \frac{1}{\left(\bar{w}-\bar{z}_2\right)^{a^{\prime}}} \frac{1}{\left(w-z_2\right)^{b^{\prime}}} \\
  &\hspace{5cm}\times\left\langle\mathcal{O}^{CD}_{\Delta_2,+1}(z_2,\bar{z}_2) \cdots \mathcal{O}^{*'_j}_{\Delta_j,\ell_j+1}(z_j,\bar{z}_j)\cdots\mathcal{O}^{*_N}_{\Delta_N,\ell_N}(z_N,\bar{z}_N) \right\rangle\\
  &=\frac{1}{\pi } \sum_{j=3}^{N}f(A,B,\ell_j,*_j,*_j')\bar{f}(C,D,\ell_j+1,*_j',*_j'')C_1(b, a)\frac{1}{\left(z_j-z\right)^{a-2}} \frac{1}{\left(\bar{z}_j-\bar{z}\right)^b}\\
  &\times\int d^2z_2 \frac{1}{\left(\bar{w}-\bar{z}_2\right)^{a^{\prime}}} \frac{1}{\left(w-z_2\right)^{b^{\prime}}}\frac{\bar{z}_2-\bar{z}_j}{z_2-z_j}\left\langle\mathcal{O}^{*_3}_{\Delta_3,\ell_3}(z_3,\bar{z}_3) \cdots \mathcal{O}^{*''_j}_{\Delta_j,\ell_j}(z_j,\bar{z}_j)\cdots\mathcal{O}^{*_N}_{\Delta_N,\ell_N}(z_N,\bar{z}_N) \right\rangle\\
  &+\frac{1}{\pi } \sum_{\substack{i,j=3\\i\neq j}}^{N}f(A,B,\ell_j,*_j,*_j')\bar{f}(C,D,\ell_i,*_i,*_i')C_1(b, a)\frac{1}{\left(z_j-z\right)^{a-2}} \frac{1}{\left(\bar{z}_j-\bar{z}\right)^b}\\
  &\times\int d^2z_2 \frac{1}{\left(\bar{w}-\bar{z}_2\right)^{a^{\prime}}} \frac{1}{\left(w-z_2\right)^{b^{\prime}}}\frac{\bar{z}_2-\bar{z}_i}{z_2-z_i}\left\langle\cdots \mathcal{O}^{*'_j}_{\Delta_j,\ell_j+1}(z_j,\bar{z}_j)\cdots\mathcal{O}^{*'_i}_{\Delta_i,\ell_i-1}(z_i,\bar{z}_i)\cdots\mathcal{O}^{*_N}_{\Delta_N,\ell_N}(z_N,\bar{z}_N) \right\rangle\\
  &=\sum_{j=3}^{N}f(A,B,\ell_j,*_j,*_j')\bar{f}(C,D,\ell_j+1,*_j',*_j'')C_1(b^{\prime}, a^{\prime})\frac{1}{\left(z_j-z\right)^{a-2}} \frac{1}{\left(\bar{z}_j-\bar{z}\right)^b}\\
  &\times\frac{1}{\left(z_j-w\right)^{b^{\prime}}} \frac{1}{\left(\bar{z}_j-\bar{w}\right)^{a^{\prime}-2}}\left\langle\mathcal{O}^{*_3}_{\Delta_3,\ell_3}(z_3,\bar{z}_3) \cdots \mathcal{O}^{*^{''}_j}_{\Delta_j,\ell_j}(z_j,\bar{z}_j)\cdots\mathcal{O}^{*_N}_{\Delta_N,\ell_N}(z_N,\bar{z}_N) \right\rangle\\
  &+ \sum_{\substack{i,j=3\\i\neq j}}^{N}f(A,B,\ell_j,*_j,*_j')\bar{f}(C,D,\ell_i,*_i,*_i')C_1(b, a)C_1(b^{\prime}, a^{\prime})\frac{1}{\left(z_j-z\right)^{a-2}} \frac{1}{\left(\bar{z}_j-\bar{z}\right)^b}\\
  &\times\frac{1}{\left(z_i-w\right)^{b^{\prime}}} \frac{1}{\left(\bar{z}_i-\bar{w}\right)^{a^{\prime}-2}}\left\langle\cdots \mathcal{O}^{*'_j}_{\Delta_j,\ell_j+1}(z_j,\bar{z}_j)\cdots\mathcal{O}^{*'_i}_{\Delta_i,\ell_i-1}(z_i,\bar{z}_i)\cdots\mathcal{O}^{*_N}_{\Delta_N,\ell_N}(z_N,\bar{z}_N) \right\rangle
\end{split}
\end{equation}
Combining the two integrals we get
\begin{equation}\label{GbarG}
    \begin{split}
       &\left\langle G_{AB}(z, \bar{z})\overline{G}^{CD}(w, \bar{w})\prod_{n=3}^N\mathcal{O}^{*_n}_{\Delta_n,\ell_n}(z_n,\bar{z}_n)\right\rangle\\
        &=-\delta^{CD}_{AB}C_1(b, a)\frac{1}{ \pi }\int d^2z_2 \frac{1}{\left(\bar{w}-\bar{z}_2\right)^{a^{\prime}}} \frac{1}{\left(w-z_2\right)^{b^{\prime}}} \frac{1}{\left(z_2-z\right)^{a-2}\left(\bar{z}_2-\bar{z}\right)^b}\\
    &\times \sum_{i=3}^{N}(1-f(\Delta_i))\frac{\bar{z}_2- \bar{z}_i}{z_2-z_i}\times\left\langle\mathcal{O}^{*_3}_{\Delta_3,\ell_3}(z_3,\bar{z}_3)\cdots\mathcal{O}^{*_i}_{\Delta_i,\ell_i}(z_i,\bar{z}_i)\cdots\mathcal{O}^{*_N}_{\Delta_N,\ell_N}(z_N,\bar{z}_N)\right\rangle\\
    &+\sum_{j=3}^{N}f(A,B,\ell_j,*_j,*_j')\bar{f}(C,D,\ell_j+1,*_j',*_j'')C_1(b^{\prime}, a^{\prime})\frac{1}{\left(z_j-z\right)^{a-2}} \frac{1}{\left(\bar{z}_j-\bar{z}\right)^b}\\
  &\times\frac{1}{\left(z_j-w\right)^{b^{\prime}}} \frac{1}{\left(\bar{z}_j-\bar{w}\right)^{a^{\prime}-2}}\left\langle\mathcal{O}^{*_3}_{\Delta_3,\ell_3}(z_3,\bar{z}_3) \cdots \mathcal{O}^{*^{''}_j}_{\Delta_j,\ell_j}(z_j,\bar{z}_j)\cdots\mathcal{O}^{*_N}_{\Delta_N,\ell_N}(z_N,\bar{z}_N) \right\rangle\\
  &+ \sum_{\substack{i,j=3\\i\neq j}}^{N}f(A,B,\ell_j,*_j,*_j')\bar{f}(C,D,\ell_i,*_i,*_i')C_1(b, a)C_1(b^{\prime}, a^{\prime})\frac{1}{\left(z_j-z\right)^{a-2}} \frac{1}{\left(\bar{z}_j-\bar{z}\right)^b}\\
  &\times\frac{1}{\left(z_i-w\right)^{b^{\prime}}} \frac{1}{\left(\bar{z}_i-\bar{w}\right)^{a^{\prime}-2}}\left\langle\cdots \mathcal{O}^{*'_j}_{\Delta_j,\ell_j+1}(z_j,\bar{z}_j)\cdots\mathcal{O}^{*'_i}_{\Delta_i,\ell_i-1}(z_i,\bar{z}_i)\cdots\mathcal{O}^{*_N}_{\Delta_N,\ell_N}(z_N,\bar{z}_N) \right\rangle
    \end{split}
\end{equation}

Here when we take the normal order of this composite current we only need to care about the non-singular terms in the above OPE. The non-singular term in the integral above can be obtained by taking $z \to w$ limit in the integral. The integral can then be evaluated as  
\[
\begin{aligned}
\int d^2 z_2 \frac{1}{\left(\bar{w}-\bar{z}_2\right)^{a^{\prime}}} &\frac{1}{\left(w-z_2\right)^{b^{\prime}}} \frac{1}{\left(z_2-z\right)^{a-2}} \frac{1}{\left(\bar{z}_2-\bar{z}\right)^b} \frac{\bar{z}_2-\bar{z}_i}{z_2-z_i} \quad \stackrel{z = w}{\longrightarrow} \\
&=(-1)^{a+b-2} \int d^2 z_2 \frac{1}{\left(\bar{z}-\bar{z}_2\right)^{a^{\prime}+b}} \frac{1}{\left(z-z_2\right)^{b^{\prime}+a-2}} \frac{\bar{z}_2-\bar{z}_i}{z_2-z_i}\\
&=(-1)^{a+b} C_1\left(a+b^{\prime}-2, a^{\prime}+b\right) \frac{1}{\left(z_i-z\right)^{a^{\prime}+b-2}} \frac{1}{\left(\bar{z}_i-\bar{z}\right)^{a+b^{\prime}-2}}
\end{aligned}
\]
Hence
\begin{equation}\label{GbarG}
    \begin{split}
     &\left\langle :G_{AB}(z, \bar{z})\overline{G}^{CD}(z,\bar{z}):\prod_{n=3}^N\mathcal{O}^{*_n}_{\Delta_n,\ell_n}(z_n,\bar{z}_n)\right\rangle\\
    &=\frac{(-1)^{a+b+1}}{ \pi }\delta^{CD}_{AB}C_1(b, a) C\left(a+b^{\prime}-2, a^{\prime}+b\right) \sum_{i=3}^{N}(1-f(\Delta_i)) \\
    &\hspace{1cm}\times\frac{1}{\left(z_i-z\right)^{a^{\prime}+b-2}}\frac{1}{\left(\bar{z}_i-\bar{z}\right)^{a+b^{\prime}-2}}\left\langle\mathcal{O}^{*_3}_{\Delta_3,\ell_3}(z_3,\bar{z}_3)\cdots\mathcal{O}^{*_i}_{\Delta_i,\ell_i}(z_i,\bar{z}_i)\cdots\mathcal{O}^{*_N}_{\Delta_N,\ell'_N}(z_N,\bar{z}_N)\right\rangle\\
    &+\sum_{j=3}^{N}f(A,B,\ell_j,*_j,*_j')\bar{f}(C,D,\ell_j+1,*_j',*_j'')C_1(b, a)C_1(b^{\prime}, a^{\prime})(-1)^{a+b+a^{\prime}+b^{\prime}}\\
  &\hspace{1cm}\times\frac{1}{\left(z-z_j\right)^{a+b^{\prime}-2}} \frac{1}{\left(\bar{z}-\bar{z}_j\right)^{a^{\prime}+b-2}}\left\langle\mathcal{O}^{*_3}_{\Delta_3,\ell_3}(z_3,\bar{z}_3) \cdots \mathcal{O}^{*^{''}_j}_{\Delta_j,\ell_j}(z_j,\bar{z}_j)\cdots\mathcal{O}^{*_N}_{\Delta_N,\ell_N}(z_N,\bar{z}_N) \right\rangle\\
  &+ \sum_{\substack{i,j=3\\i\neq j}}^{N}f(A,B,\ell_j,*_j,*_j')\bar{f}(C,D,\ell_i,*_i,*_i')C_1(b, a)C_1(b^{\prime}, a^{\prime})(-1)^{a+b+a^{\prime}+b^{\prime}}\frac{1}{\left(z-z_j\right)^{a-2}} \frac{1}{\left(\bar{z}-\bar{z}_j\right)^{b}}\\
  &\hspace{1cm}\times\frac{1}{\left(z-z_i\right)^{b^{\prime}}} \frac{1}{\left(\bar{z}-\bar{z}_i\right)^{a^{\prime}-2}}\left\langle\cdots \mathcal{O}^{*'_j}_{\Delta_j,\ell_j+1}(z_j,\bar{z}_j)\cdots\mathcal{O}^{*'_i}_{\Delta_i,\ell_i-1}(z_i,\bar{z}_i)\cdots\mathcal{O}^{*_N}_{\Delta_N,\ell_N}(z_N,\bar{z}_N) \right\rangle
    \end{split}
\end{equation}
Similarly we have,
\begin{equation}\label{barGG}
    \begin{split}
        &\left\langle :\overline{G}^{CD}(z,\bar{z})G_{AB}(z,\bar{z}):\prod_{n=3}^N\mathcal{O}^{*_n}_{\Delta_n,\ell_n}(z_n,\bar{z}_n)\right\rangle\\
        &=\frac{(-1)^{a^{\prime}+b^{\prime}+1}}{ \pi }\delta^{CD}_{AB}C_1(b^{\prime}, a^{\prime}) C\left(a^{\prime}+b-2, a+b^{\prime}\right)\sum_{i=3}^{N}(1-f(\Delta_i)) \\
    &\hspace{1cm}\times\frac{1}{\left(z_i-z\right)^{a+b^{\prime}-2}} \frac{1}{\left(\bar{z}_i-\bar{z}\right)^{a^{\prime}+b-2}}\left\langle\mathcal{O}^{*_3}_{\Delta_3,\ell_3}(z_3,\bar{z}_3)\cdots\mathcal{O}^{*_i}_{\Delta_i,\ell_i}(z_i,\bar{z}_i)\cdots\mathcal{O}^{*_N}_{\Delta_N,\ell_N}(z_N,\bar{z}_N)\right\rangle\\
    &+\sum_{j=3}^{N}\bar{f}(C,D,\ell_j,*_j,*_j')f(A,B,\ell_j-1,*_j',*_j'')C_1(b^{\prime}, a^{\prime})C_1(b, a)(-1)^{a+b+a^{\prime}+b^{\prime}}\\
  &\hspace{1cm}\times\frac{1}{\left(z-z_j\right)^{a^{\prime}+b-2}} \frac{1}{\left(\bar{z}-\bar{z}_j\right)^{a+b^{\prime}-2}}\left\langle\mathcal{O}^{*_3}_{\Delta_3,\ell_3}(z_3,\bar{z}_3) \cdots \mathcal{O}^{*^{''}_j}_{\Delta_j,\ell_j}(z_j,\bar{z}_j)\cdots\mathcal{O}^{*_N}_{\Delta_N,\ell_N}(z_N,\bar{z}_N) \right\rangle\\
  &+ \sum_{\substack{i,j=3\\i\neq j}}^{N}\bar{f}(C,D,\ell_j,*_j,*_j')f(A,B,\ell_i,*_i,*_i')C_1(b, a)C_1(b^{\prime}, a^{\prime})(-1)^{a+b+a^{\prime}+b^{\prime}}\frac{1}{\left(\bar{z}-\bar{z}_j\right)^{a^{\prime}-2}} \frac{1}{\left(z-z_j\right)^{b^{\prime}}}\\
  &\hspace{1cm}\times\frac{1}{\left(\bar{z}-\bar{z}_i\right)^{b}} \frac{1}{\left(z-z_i\right)^{a-2}}\left\langle\cdots \mathcal{O}^{*'_j}_{\Delta_j,\ell_j-1}(z_j,\bar{z}_j)\cdots\mathcal{O}^{*'_i}_{\Delta_i,\ell_i+1}(z_i,\bar{z}_i)\cdots\mathcal{O}^{*_N}_{\Delta_N,\ell_N}(z_N,\bar{z}_N) \right\rangle
    \end{split}
\end{equation}
We have the correlator of the normalised current with any conformal primary as,
\begin{equation}
    \begin{split}
&\left\langle \mathcal{G}_{AB}^{CD}(z, \bar{z} )\prod_{n=3}^{N}\mathcal{O}^{*_n}_{\Delta_n,\ell_n}(z_n,\bar{z}_n)\right\rangle\\
&= -\delta^{CD}_{AB}\sum_{i=3}^{N}(1-f(\Delta_i))\Bigg[ \frac{(-1)^{a+b}}{ \pi }C_1(b, a) C\left(a+b^{\prime}-2, a^{\prime}+b\right) \frac{1}{\left(z_i-z\right)^{a^{\prime}+b-2}} \frac{1}{\left(\bar{z}_i-\bar{z}\right)^{a+b^{\prime}-2}}\\
    &\hspace{5cm}\times\left\langle\mathcal{O}^{*_3}_{\Delta_3,\ell_3}(z_3,\bar{z}_3)\cdots\mathcal{O}^{*_i}_{\Delta_i,\ell_i}(z_i,\bar{z}_i)\cdots\mathcal{O}^{*_N}_{\Delta_N,\ell_N}(z_N,\bar{z}_N)\right\rangle\\
    &-\frac{(-1)^{a^{\prime}+b^{\prime}}}{ \pi }C_1(b^{\prime}, a^{\prime}) C_1\left(a^{\prime}+b-2, a+b^{\prime}\right) \frac{1}{\left(z_i-z\right)^{a+b^{\prime}-2}} \frac{1}{\left(\bar{z}_i-\bar{z}\right)^{a^{\prime}+b-2}}\\
    &\hspace{5cm}\times\langle\mathcal{O}^{*_3}_{\Delta_3,\ell_3}(z_3,\bar{z}_3)\cdots\mathcal{O}^{*_i}_{\Delta_i,\ell_i}(z_i,\bar{z}_i)\cdots\mathcal{O}^{*_N}_{\Delta_N,\ell_N}(z_N,\bar{z}_N)\rangle\Bigg]\\
    &+(-1)^{a+b+a^{\prime}+b^{\prime}}C_1(b, a)C_1(b^{\prime}, a^{\prime})\Bigg[f(A,B,\ell_j,*_j,*_j')\bar{f}(C,D,\ell_j+1,*_j',*_j'')\\
  &\hspace{1cm}\times\frac{1}{\left(z-z_j\right)^{a+b^{\prime}-2}} \frac{1}{\left(\bar{z}-\bar{z}_j\right)^{a^{\prime}+b-2}}\left\langle\mathcal{O}^{*_3}_{\Delta_3,\ell_3}(z_3,\bar{z}_3) \cdots \mathcal{O}^{*^{''}_j}_{\Delta_j,\ell_j}(z_j,\bar{z}_j)\cdots\mathcal{O}^{*_N}_{\Delta_N,\ell_N}(z_N,\bar{z}_N) \right\rangle\\
  &\hspace{2cm}-\bar{f}(C,D,\ell_j,*_j,*_j')f(A,B,\ell_j-1,*_j',*_j'')\\
  &\hspace{1cm}\times\frac{1}{\left(z-z_j\right)^{a^{\prime}+b-2}} \frac{1}{\left(\bar{z}-\bar{z}_j\right)^{a+b^{\prime}-2}}\left\langle\mathcal{O}^{*_3}_{\Delta_3,\ell_3}(z_3,\bar{z}_3) \cdots \mathcal{O}^{*^{''}_j}_{\Delta_j,\ell_j}(z_j,\bar{z}_j)\cdots\mathcal{O}^{*_N}_{\Delta_N,\ell_N}(z_N,\bar{z}_N) \right\rangle\Bigg]
    \end{split}
\end{equation}
The last term in both  \eqref{GbarG} and  \eqref{barGG} cancels when we take the commutator. Now in the above OPE we can see that the first term which has the graviton soft limits, does not satisfy our requirement explained in \eqref{eq:curGreqopeSSbar}. Hence we require that the first terms in the two ordering to be the same, so that they cancel once we take the commutator. This is equivalent to the requirement
\begin{equation}\label{state1}
   \begin{aligned}
C_1\left(b^{\prime}, a^{\prime}\right) C_1\left(a^{\prime}+b-2, a+b^{\prime}\right)(-1)^{a+b} &=C_1(b, a) C_1\left(a+b^{\prime}-2, a^{\prime}+b\right)(-1)^{a^{\prime}+b^{\prime}} 
\end{aligned} 
\end{equation}
and 
\begin{equation}\label{state}
\begin{split}
    a^{\prime}+b-2 &=a+b^{\prime}-2
\end{split}
\end{equation}
Now in  \eqref{state1} by substituting the explicit expression from \eqref{st} we have
\[
\left(-a^{\prime}-b+1\right)\left(-a^{\prime}-b+2\right)(-a+1)(-a+2)=\left(-a^{\prime}+1\right)\left(-a^{\prime}+2\right)\left(-a-b^{\prime}+1\right)\left(-a-b^{\prime}+2\right).
\]
which after using \eqref{state} gives 
\begin{equation}
(-a+1)(-a+2)=\left(-a^{\prime}+1\right)\left(-a^{\prime}+2\right)    
\end{equation}
which clearly has solutions.
Hence correlator corresponding to this normal order current is,
\begin{equation}\label{compon}
    \begin{split}
&\left\langle \mathcal{G}_{AB}^{CD}(z, \bar{z} )\prod_{n=3}^{N}\mathcal{O}^{*_n}_{\Delta_n,\ell_n}(z_n,\bar{z}_n)\right\rangle\\
&=  (-1)^{a+b+a^{\prime}+b^{\prime}}C_1(b, a)C_1(b^{\prime}, a^{\prime})\Bigg[f(A,B,\ell_j,*_j,*_j')\bar{f}(C,D,\ell_j+1,*_j',*_j'')\\
  &\hspace{1cm}\times\frac{1}{\left(z-z_j\right)^{a+b^{\prime}-2}} \frac{1}{\left(\bar{z}-\bar{z}_j\right)^{a^{\prime}+b-2}}\left\langle\mathcal{O}^{*_3}_{\Delta_3,\ell_3}(z_3,\bar{z}_3) \cdots \mathcal{O}^{*^{''}_j}_{\Delta_j,\ell_j}(z_j,\bar{z}_j)\cdots\mathcal{O}^{*_N}_{\Delta_N,\ell_N}(z_N,\bar{z}_N) \right\rangle\\
  &\hspace{2cm}-\bar{f}(C,D,\ell_j,*_j,*_j')f(A,B,\ell_j-1,*_j',*_j'')\\
  &\hspace{1cm}\times\frac{1}{\left(z-z_j\right)^{a^{\prime}+b-2}} \frac{1}{\left(\bar{z}-\bar{z}_j\right)^{a+b^{\prime}-2}}\left\langle\mathcal{O}^{*_3}_{\Delta_3,\ell_3}(z_3,\bar{z}_3) \cdots \mathcal{O}^{*^{''}_j}_{\Delta_j,\ell_j}(z_j,\bar{z}_j)\cdots\mathcal{O}^{*_N}_{\Delta_N,\ell_N}(z_N,\bar{z}_N) \right\rangle\Bigg]
\end{split}
\end{equation}
\end{appendix}
\bibliography{sbms.bib}

\providecommand{\href}[2]{#2}\begingroup\raggedright\begin{thebibliography}{10}

\bibitem{Fotopoulos:2020bqj}
A.~Fotopoulos, S.~Stieberger, T.~R. Taylor and B.~Zhu, \emph{{Extended Super
  BMS Algebra of Celestial CFT}},
  \href{https://doi.org/10.1007/JHEP09(2020)198}{\emph{JHEP} {\bfseries 09}
  (2020) 198} [\href{https://arxiv.org/abs/2007.03785}{{\ttfamily
  2007.03785}}].

\bibitem{He:2014cra}
T.~He, P.~Mitra, A.~P. Porfyriadis and A.~Strominger, \emph{{New Symmetries of
  Massless QED}}, \href{https://doi.org/10.1007/JHEP10(2014)112}{\emph{JHEP}
  {\bfseries 10} (2014) 112} [\href{https://arxiv.org/abs/1407.3789}{{\ttfamily
  1407.3789}}].

\bibitem{He:2015zea}
T.~He, P.~Mitra and A.~Strominger, \emph{{2D Kac-Moody Symmetry of 4D
  Yang-Mills Theory}},
  \href{https://doi.org/10.1007/JHEP10(2016)137}{\emph{JHEP} {\bfseries 10}
  (2016) 137} [\href{https://arxiv.org/abs/1503.02663}{{\ttfamily
  1503.02663}}].

\bibitem{Dumitrescu:2015fej}
T.~T. Dumitrescu, T.~He, P.~Mitra and A.~Strominger,
  \emph{{Infinite-dimensional fermionic symmetry in supersymmetric gauge
  theories}}, \href{https://doi.org/10.1007/JHEP08(2021)051}{\emph{JHEP}
  {\bfseries 08} (2021) 051}
  [\href{https://arxiv.org/abs/1511.07429}{{\ttfamily 1511.07429}}].

\bibitem{PhysRevLett.105.111103}
G.~Barnich and C.~Troessaert, \emph{Symmetries of asymptotically flat
  four-dimensional spacetimes at null infinity revisited},
  \href{https://doi.org/10.1103/PhysRevLett.105.111103}{\emph{Phys. Rev. Lett.}
  {\bfseries 105} (2010) 111103}.

\bibitem{PhysRev.128.2851}
R.~Sachs, \emph{Asymptotic symmetries in gravitational theory},
  \href{https://doi.org/10.1103/PhysRev.128.2851}{\emph{Phys. Rev.} {\bfseries
  128} (1962) 2851}.

\bibitem{Strominger:2013lka}
A.~Strominger, \emph{{Asymptotic Symmetries of Yang-Mills Theory}},
  \href{https://doi.org/10.1007/JHEP07(2014)151}{\emph{JHEP} {\bfseries 07}
  (2014) 151} [\href{https://arxiv.org/abs/1308.0589}{{\ttfamily 1308.0589}}].

\bibitem{Bondi:1962px}
H.~Bondi, M.~G.~J. van~der Burg and A.~W.~K. Metzner, \emph{{Gravitational
  waves in general relativity. 7. Waves from axisymmetric isolated systems}},
  \href{https://doi.org/10.1098/rspa.1962.0161}{\emph{Proc. Roy. Soc. Lond. A}
  {\bfseries 269} (1962) 21}.

\bibitem{Sachs:1962wk}
R.~K. Sachs, \emph{{Gravitational waves in general relativity. 8. Waves in
  asymptotically flat space-times}},
  \href{https://doi.org/10.1098/rspa.1962.0206}{\emph{Proc. Roy. Soc. Lond. A}
  {\bfseries 270} (1962) 103}.

\bibitem{Sachs:1962zza}
R.~Sachs, \emph{{Asymptotic symmetries in gravitational theory}},
  \href{https://doi.org/10.1103/PhysRev.128.2851}{\emph{Phys. Rev.} {\bfseries
  128} (1962) 2851}.

\bibitem{Strominger:2017zoo}
A.~Strominger, \emph{{Lectures on the Infrared Structure of Gravity and Gauge
  Theory}},  \href{https://arxiv.org/abs/1703.05448}{{\ttfamily 1703.05448}}.

\bibitem{He:2014laa}
T.~He, V.~Lysov, P.~Mitra and A.~Strominger, \emph{{BMS supertranslations and
  Weinberg\textquoteright{}s soft graviton theorem}},
  \href{https://doi.org/10.1007/JHEP05(2015)151}{\emph{JHEP} {\bfseries 05}
  (2015) 151} [\href{https://arxiv.org/abs/1401.7026}{{\ttfamily 1401.7026}}].

\bibitem{Barnich:2011mi}
G.~Barnich and C.~Troessaert, \emph{{BMS charge algebra}},
  \href{https://doi.org/10.1007/JHEP12(2011)105}{\emph{JHEP} {\bfseries 12}
  (2011) 105} [\href{https://arxiv.org/abs/1106.0213}{{\ttfamily 1106.0213}}].

\bibitem{Barnich:2013axa}
G.~Barnich and C.~Troessaert, \emph{{Comments on holographic current algebras
  and asymptotically flat four dimensional spacetimes at null infinity}},
  \href{https://doi.org/10.1007/JHEP11(2013)003}{\emph{JHEP} {\bfseries 11}
  (2013) 003} [\href{https://arxiv.org/abs/1309.0794}{{\ttfamily 1309.0794}}].

\bibitem{Strominger:2013jfa}
A.~Strominger, \emph{{On BMS Invariance of Gravitational Scattering}},
  \href{https://doi.org/10.1007/JHEP07(2014)152}{\emph{JHEP} {\bfseries 07}
  (2014) 152} [\href{https://arxiv.org/abs/1312.2229}{{\ttfamily 1312.2229}}].

\bibitem{Gabai:2016kuf}
B.~Gabai and A.~Sever, \emph{{Large gauge symmetries and asymptotic states in
  QED}}, \href{https://doi.org/10.1007/JHEP12(2016)095}{\emph{JHEP} {\bfseries
  12} (2016) 095} [\href{https://arxiv.org/abs/1607.08599}{{\ttfamily
  1607.08599}}].

\bibitem{Favata:2010zu}
M.~Favata, \emph{{The gravitational-wave memory effect}},
  \href{https://doi.org/10.1088/0264-9381/27/8/084036}{\emph{Class. Quant.
  Grav.} {\bfseries 27} (2010) 084036}
  [\href{https://arxiv.org/abs/1003.3486}{{\ttfamily 1003.3486}}].

\bibitem{PhysRevLett.67.1486}
D.~Christodoulou, \emph{Nonlinear nature of gravitation and gravitational-wave
  experiments}, \href{https://doi.org/10.1103/PhysRevLett.67.1486}{\emph{Phys.
  Rev. Lett.} {\bfseries 67} (1991) 1486}.

\bibitem{Zeldovich:1974gvh}
Y.~B. Zel'dovich and A.~G. Polnarev, \emph{{Radiation of gravitational waves by
  a cluster of superdense stars}}, {\emph{Sov. Astron.} {\bfseries 18} (1974)
  17}.

\bibitem{Susskind:2015hpa}
L.~Susskind, \emph{{Electromagnetic Memory}},
  \href{https://arxiv.org/abs/1507.02584}{{\ttfamily 1507.02584}}.

\bibitem{Pasterski:2015zua}
S.~Pasterski, \emph{{Asymptotic Symmetries and Electromagnetic Memory}},
  \href{https://doi.org/10.1007/JHEP09(2017)154}{\emph{JHEP} {\bfseries 09}
  (2017) 154} [\href{https://arxiv.org/abs/1505.00716}{{\ttfamily
  1505.00716}}].

\bibitem{Carlip:2017xne}
S.~Carlip, \emph{{Black Hole Entropy from Bondi-Metzner-Sachs Symmetry at the
  Horizon}}, \href{https://doi.org/10.1103/PhysRevLett.120.101301}{\emph{Phys.
  Rev. Lett.} {\bfseries 120} (2018) 101301}
  [\href{https://arxiv.org/abs/1702.04439}{{\ttfamily 1702.04439}}].

\bibitem{Sen:2007qy}
A.~Sen, \emph{{Black Hole Entropy Function, Attractors and Precision Counting
  of Microstates}}, \href{https://doi.org/10.1007/s10714-008-0626-4}{\emph{Gen.
  Rel. Grav.} {\bfseries 40} (2008) 2249}
  [\href{https://arxiv.org/abs/0708.1270}{{\ttfamily 0708.1270}}].

\bibitem{Mandal:2010cj}
I.~Mandal and A.~Sen, \emph{{Black Hole Microstate Counting and its Macroscopic
  Counterpart}},
  \href{https://doi.org/10.1088/0264-9381/27/21/214003}{\emph{Class. Quant.
  Grav.} {\bfseries 27} (2010) 214003}
  [\href{https://arxiv.org/abs/1008.3801}{{\ttfamily 1008.3801}}].

\bibitem{Dabholkar:2004yr}
A.~Dabholkar, \emph{{Exact counting of black hole microstates}},
  \href{https://doi.org/10.1103/PhysRevLett.94.241301}{\emph{Phys. Rev. Lett.}
  {\bfseries 94} (2005) 241301}
  [\href{https://arxiv.org/abs/hep-th/0409148}{{\ttfamily hep-th/0409148}}].

\bibitem{Carlip:2012ff}
S.~Carlip, \emph{{Effective Conformal Descriptions of Black Hole Entropy: A
  Review}}, \href{https://doi.org/10.1063/1.4756962}{\emph{AIP Conf. Proc.}
  {\bfseries 1483} (2012) 54}
  [\href{https://arxiv.org/abs/1207.1488}{{\ttfamily 1207.1488}}].

\bibitem{Campiglia:2019wxe}
M.~Campiglia and A.~Laddha, \emph{{Loop Corrected Soft Photon Theorem as a Ward
  Identity}}, \href{https://doi.org/10.1007/JHEP10(2019)287}{\emph{JHEP}
  {\bfseries 10} (2019) 287}
  [\href{https://arxiv.org/abs/1903.09133}{{\ttfamily 1903.09133}}].

\bibitem{Taylor:2017sph}
T.~R. Taylor, \emph{{A Course in Amplitudes}},
  \href{https://doi.org/10.1016/j.physrep.2017.05.002}{\emph{Phys. Rept.}
  {\bfseries 691} (2017) 1} [\href{https://arxiv.org/abs/1703.05670}{{\ttfamily
  1703.05670}}].

\bibitem{Pasterski:2021raf}
S.~Pasterski, M.~Pate and A.-M. Raclariu, \emph{{Celestial Holography}},  in
  \emph{{2022 Snowmass Summer Study}}, 11, 2021,
  \href{https://arxiv.org/abs/2111.11392}{{\ttfamily 2111.11392}}.

\bibitem{Pasterski:2021rjz}
S.~Pasterski, \emph{{Lectures on celestial amplitudes}},
  \href{https://doi.org/10.1140/epjc/s10052-021-09846-7}{\emph{Eur. Phys. J. C}
  {\bfseries 81} (2021) 1062}
  [\href{https://arxiv.org/abs/2108.04801}{{\ttfamily 2108.04801}}].

\bibitem{Donnay:2020guq}
L.~Donnay, S.~Pasterski and A.~Puhm, \emph{{Asymptotic Symmetries and Celestial
  CFT}}, \href{https://doi.org/10.1007/JHEP09(2020)176}{\emph{JHEP} {\bfseries
  09} (2020) 176} [\href{https://arxiv.org/abs/2005.08990}{{\ttfamily
  2005.08990}}].

\bibitem{Fan:2019emx}
W.~Fan, A.~Fotopoulos and T.~R. Taylor, \emph{{Soft Limits of Yang-Mills
  Amplitudes and Conformal Correlators}},
  \href{https://doi.org/10.1007/JHEP05(2019)121}{\emph{JHEP} {\bfseries 05}
  (2019) 121} [\href{https://arxiv.org/abs/1903.01676}{{\ttfamily
  1903.01676}}].

\bibitem{Fotopoulos:2019vac}
A.~Fotopoulos, S.~Stieberger, T.~R. Taylor and B.~Zhu, \emph{{Extended BMS
  Algebra of Celestial CFT}},
  \href{https://doi.org/10.1007/JHEP03(2020)130}{\emph{JHEP} {\bfseries 03}
  (2020) 130} [\href{https://arxiv.org/abs/1912.10973}{{\ttfamily
  1912.10973}}].

\bibitem{Banerjee:2022wht}
S.~Banerjee and S.~Pasterski, \emph{{Revisiting the Shadow Stress Tensor in
  Celestial CFT}},  \href{https://arxiv.org/abs/2212.00257}{{\ttfamily
  2212.00257}}.

\bibitem{Banerjee:2021uxe}
N.~Banerjee, T.~Rahnuma and R.~K. Singh, \emph{{Asymptotic Symmetry of Four
  Dimensional Einstein-Yang-Mills and Einstein-Maxwell Theory}},
  \href{https://arxiv.org/abs/2110.15657}{{\ttfamily 2110.15657}}.

\bibitem{Banerjee:2022abf}
N.~Banerjee, A.~Mitra, D.~Mukherjee and H.~R. Safari,
  \emph{{Supersymmetrization of deformed BMS algebras}},
  \href{https://arxiv.org/abs/2201.09853}{{\ttfamily 2201.09853}}.

\bibitem{tab}
N.~Banerjee, T.~Rahnuma and R.~K. Singh, \emph{{Soft and Collinear Limits in
  $\mathcal{N}=8$ Supergravity using Double Copy Formalism}},
  \href{https://arxiv.org/abs/2212.11480}{{\ttfamily 2212.11480}}.

\bibitem{Jiang:2021xzy}
H.~Jiang, \emph{{Celestial superamplitude in $ \mathcal{N} $ = 4 SYM theory}},
  \href{https://doi.org/10.1007/JHEP08(2021)031}{\emph{JHEP} {\bfseries 08}
  (2021) 031} [\href{https://arxiv.org/abs/2105.10269}{{\ttfamily
  2105.10269}}].

\bibitem{He:2014bga}
S.~He, Y.-t. Huang and C.~Wen, \emph{{Loop Corrections to Soft Theorems in
  Gauge Theories and Gravity}},
  \href{https://doi.org/10.1007/JHEP12(2014)115}{\emph{JHEP} {\bfseries 12}
  (2014) 115} [\href{https://arxiv.org/abs/1405.1410}{{\ttfamily 1405.1410}}].

\bibitem{Adamo:2019ipt}
T.~Adamo, L.~Mason and A.~Sharma, \emph{{Celestial amplitudes and conformal
  soft theorems}}, \href{https://doi.org/10.1088/1361-6382/ab42ce}{\emph{Class.
  Quant. Grav.} {\bfseries 36} (2019) 205018}
  [\href{https://arxiv.org/abs/1905.09224}{{\ttfamily 1905.09224}}].

\bibitem{Bianchi:2008pu}
M.~Bianchi, H.~Elvang and D.~Z. Freedman, \emph{{Generating Tree Amplitudes in
  N=4 SYM and N = 8 SG}},
  \href{https://doi.org/10.1088/1126-6708/2008/09/063}{\emph{JHEP} {\bfseries
  09} (2008) 063} [\href{https://arxiv.org/abs/0805.0757}{{\ttfamily
  0805.0757}}].

\bibitem{Fan:2020xjj}
W.~Fan, A.~Fotopoulos, S.~Stieberger and T.~R. Taylor, \emph{{On Sugawara
  construction on Celestial Sphere}},
  \href{https://doi.org/10.1007/JHEP09(2020)139}{\emph{JHEP} {\bfseries 09}
  (2020) 139} [\href{https://arxiv.org/abs/2005.10666}{{\ttfamily
  2005.10666}}].

\bibitem{Jiang:2021ovh}
H.~Jiang, \emph{{Holographic chiral algebra: supersymmetry, infinite Ward
  identities, and EFTs}},
  \href{https://doi.org/10.1007/JHEP01(2022)113}{\emph{JHEP} {\bfseries 01}
  (2022) 113} [\href{https://arxiv.org/abs/2108.08799}{{\ttfamily
  2108.08799}}].

\bibitem{Hawking:2016sgy}
S.~W. Hawking, M.~J. Perry and A.~Strominger, \emph{{Superrotation Charge and
  Supertranslation Hair on Black Holes}},
  \href{https://doi.org/10.1007/JHEP05(2017)161}{\emph{JHEP} {\bfseries 05}
  (2017) 161} [\href{https://arxiv.org/abs/1611.09175}{{\ttfamily
  1611.09175}}].

\bibitem{Haco:2018ske}
S.~Haco, S.~W. Hawking, M.~J. Perry and A.~Strominger, \emph{{Black Hole
  Entropy and Soft Hair}},
  \href{https://doi.org/10.1007/JHEP12(2018)098}{\emph{JHEP} {\bfseries 12}
  (2018) 098} [\href{https://arxiv.org/abs/1810.01847}{{\ttfamily
  1810.01847}}].

\bibitem{Hawking:2016msc}
S.~W. Hawking, M.~J. Perry and A.~Strominger, \emph{{Soft Hair on Black
  Holes}}, \href{https://doi.org/10.1103/PhysRevLett.116.231301}{\emph{Phys.
  Rev. Lett.} {\bfseries 116} (2016) 231301}
  [\href{https://arxiv.org/abs/1601.00921}{{\ttfamily 1601.00921}}].

\bibitem{Chu:2018tzu}
C.-S. Chu and Y.~Koyama, \emph{{Soft Hair of Dynamical Black Hole and Hawking
  Radiation}}, \href{https://doi.org/10.1007/JHEP04(2018)056}{\emph{JHEP}
  {\bfseries 04} (2018) 056}
  [\href{https://arxiv.org/abs/1801.03658}{{\ttfamily 1801.03658}}].

\bibitem{Averin:2016hhm}
A.~Averin, G.~Dvali, C.~Gomez and D.~Lust, \emph{{Goldstone origin of black
  hole hair from supertranslations and criticality}},
  \href{https://doi.org/10.1142/S0217732316300457}{\emph{Mod. Phys. Lett. A}
  {\bfseries 31} (2016) 1630045}
  [\href{https://arxiv.org/abs/1606.06260}{{\ttfamily 1606.06260}}].

\end{thebibliography}\endgroup
\end{document}